 \documentclass[5p,times,authoryear]{elsarticle}

\usepackage{amssymb}
\usepackage{lipsum}
\usepackage{amsmath}
\usepackage{float}
\usepackage{threeparttable} 
\usepackage{multirow}
\usepackage{diagbox}
\usepackage{makecell}
\usepackage[T1]{fontenc}	

\usepackage{comment}

\usepackage{hyperref}
\hypersetup{
    colorlinks=true,
    linkcolor=blue,
    filecolor=magenta,      
    urlcolor=cyan,
    pdftitle={RFI-DRUnet: Restoring dynamic spectra corrupted by radio frequency interference~-- Application to pulsar observations},
    }

\newcommand{\indFreq}{n}
\newcommand{\indTime}{k}
\newcommand{\NFreq}{N}
\newcommand{\NTime}{K}

\journal{Submitted to a journal}

\begin{document}

\begin{frontmatter}

\title{RFI-DRUnet: Restoring dynamic spectra corrupted by radio frequency {interference}~-- \\Application to pulsar observations}

\tnotetext[]{Part of this work has been supported by the ANR-3IA Artificial, Natural Intelligence Toulouse Institute (ANITI) under grant agreement ANITI ANR-19-P3IA-0004 and by the CNRS through the 80|PRIME Program.}

\author[lpc,irit]{Xiao Zhang}
\ead{xiao.zhang@cnrs-orleans.fr}
\author[lpc,nancay]{Isma\"el Cognard}
\ead{icognard@cnrs-orleans.fr}
\author[irit]{Nicolas Dobigeon}
\ead{nicolas.dobigeon@enseeiht.fr}

\affiliation[lpc]{organization={Laboratoire de Physique et Chimie de l'Environnement et de l'Espace (LPC2E), CNRS, Universit\'e d’Orl\'eans},
            city={Orl\'eans},
            postcode={45071}, 
            country={France}}
            
\affiliation[nancay]{organization={Station de Radioastronomie de Nan\c{c}ay, Observatoire de Paris, PSL Research University, CNRS/INSU},
            city={Nan\c{c}ay},
            postcode={18330}, 
            country={France}}

\affiliation[irit]{organization={Institut de Recherche en Informatique de Toulouse (IRIT),  Toulouse INP, CNRS, University of Toulouse},
            city={Toulouse},
            postcode={31000}, 
            country={France}}

\begin{abstract}
Radio frequency interference (RFI) have been an enduring concern in radio astronomy, particularly for the observations of pulsars which require high timing precision and data sensitivity. In most works of the literature, RFI mitigation has been formulated as a detection task that consists of localizing possible RFI in dynamic spectra. This strategy inevitably leads to a potential loss of information since parts of the signal identified as possibly RFI-corrupted are generally not considered in the subsequent data processing pipeline. Conversely, this work proposes to tackle  RFI mitigation as a joint detection and restoration that allows parts of the dynamic spectrum affected by RFI to be not only identified but also recovered. The proposed {supervised} method relies on a deep convolutional network whose architecture inherits the performance reached by a recent yet popular image-denoising network. To train this network, a whole simulation framework is built to generate large data sets according to physics-inspired and statistical models of the pulsar signals and of the RFI. The relevance of the proposed approach is quantitatively assessed by conducting extensive experiments. In particular, the results show that the restored dynamic spectra are sufficiently reliable to estimate pulsar times-of-arrivals with an accuracy close to the one that would be obtained from RFI-free signals.

\end{abstract}


\end{frontmatter}

\section{Introduction}
\label{introduction}

Endpoint of a massive star evolution, a pulsar is a highly magnetized, rapidly rotating neutron star, which emits beams of radiation. Their primary interest lies in the fact that it is an extremely dense and compact object with a remarkably regular rotation period.
Large decimetric radio telescopes are used to study pulsars and to time the most stable ones.
This intrinsic extreme regularity can then be used to test gravitational 
theories \citep{Kramer2021} or detect very low frequency gravitational
waves \citep{Agazie2023,Antoniadis2023,Reardon2023}.

Low frequency telescopes, such as the dutch Low-Frequency Array (LOFAR) in the Netherlands \citep{stappers2011observing} or the New extension in Nan\c cay upgrading LOFAR (NenuFAR) \citep{bondonneau2021pulsars}  can be used to extend the pulsar population or to better understand the effects of interstellar propagation, critical to get a reliable timing.

In the realm of pulsar observations, especially at low frequency, radio frequency interference (RFI) pose a formidable challenge since they may significantly degrade the quality of astronomical data. They are emitted by man-made sources such as cell phones, Wi-Fi, communication satellites, and radar systems. This source diversity is reflected in the temporal and spectral ranges of the measurements that may be affected by {interference}, which makes their handling complex. Besides, RFI signals typically exhibit higher amplitudes and distinct distributions compared to astronomical signals.

These RFI corruptions can dramatically impair the study of celestial objects, especially pulsar timing which requires extremely high precision and sensitivity of the data. Thus the problem of RFI mitigation has already received considerable attention in the literature. Researchers have explored a variety of approaches, including traditional thresholding-based methods and more recent data-driven techniques capitalizing on recent advances in machine (deep) learning. However, existing RFI mitigation approaches face two main challenges. The first obstacle is a lack of ground truth, i.e., the absence of properly labeled data required to train models within supervised learning. Since manually labeling large data sets is infeasible, thresholding-based methods \citep{lazarus2016prospects} are generally employed to identify RFI in real measurements, which is then considered as the labeled ground truth. However, following such a naive strategy, the most advanced RFI mitigation techniques will certainly not be able to perform better than the unsupervised methods resorted to building the training sets, as already highlighted by \cite{Berthereau2023phd}. The second issue is the subsequent loss of information imposed by most of the existing methods, mainly due to the way the mitigation task is generally formulated. Indeed, conventional thresholding-based and data-driven RFI mitigation methods cast RFI mitigation as a binary classification or segmentation problem. In other words, they only aim at identifying and labeling parts of the dynamic spectra corrupted by RFI, also referred to as RFI flagging. However, once the parts of the dynamic spectrum corrupted by RFI have been identified, they are generally discarded from any subsequent analysis in the posterior data processing pipeline. This leads to a possibly significant loss of information, depending on the spectral and temporal ranges of the measurements affected by {interference}.

{This paper takes up the challenge of overcoming the two aforementioned issues. It proposes a new supervised RFI mitigation method preceded by a dedicated data generation framework that can be used to train this model. Indeed, to compensate for the difficulty of having access to accurately labeled ground truth data, this work first introduces a general framework to easily generate realistic dynamic spectra corrupted by RFI.} Even if previous works have considered similar strategies of data generation, none of them have been specifically designed for pulsar observations \citep{akeret2017hide,asad2021primary,deboer2017hydrogen}. This framework leverages a model-based generation of pulsar and RFI signals separately. Thanks to its versatility, this framework can be instantiated to produce realistic measurements that would have been made in various observational setups. In particular, this paper implements this general framework to produce large data sets that mimic observations by the NenuFAR telescope. 

{These data sets are subsequently used to train, validate and test a new dedicated RFI mitigation method within a supervised framework. To go beyond simple RFI tagging, this paper draws a parallel between RFI mitigation and a ubiquitous task encountered in image processing, namely image denoising. Contrary to most of the alternatives proposed in the literature, this method formulates this task as a supervised restoration problem. Benefiting from the generation framework previously introduced, the proposed approach has the ambition not only to detect and remove the RFI from the dynamic spectra but also to recover plausible signal values in place of the corrupted ones. The model is chosen as a deep neural network whose architecture inherits the performance reached by a recent yet popular image-denoising model  \citep{zhang2021plug}.}

The contributions reported in this paper can be summarized as follows: \emph{i)} the problem of RFI mitigation is envisioned as a joint detection and restoration task, which opens the door to capitalize on a recently proposed denoising-oriented deep neural architecture, \emph{ii)} to overcome the difficulty of accessing to large labeled training  data set, a versatile framework to generate RFI-corrupted dynamic spectra is proposed and instantiated to produce simulated signals compatible with real-world pulsar observations performed by NenuFAR, \emph{iii)} once trained on data sets generated following the aforementioned simulation protocol, the proposed deep network coined as {\emph{RFI denoising residual U-net} (RFI-DRUNet)} is tested through an extensive set of numerical experiments to quantitatively assess its performance and \emph{iv)} one shows that the adopted strategy allows pulsar time-of-arrival estimation to be efficiently conducted on dynamic spectra restored by RFI-DRUNet, reaching an accuracy close to the one obtained on RFI-free signals.

The remaining of the paper is organized as follows. Existing RFI mitigation methods are reviewed and discussed in Section \ref{Related Work}.  Section \ref{simulation framework} presents a versatile framework to generate RFI-corrupted pulsar observation data. The problem of RFI mitigation is envisioned from a restoration perspective in Section \ref{Deep denoising network}. The architecture of the proposed deep neural network designed to restore dynamic spectra is also detailed. Section \ref{Experiments framework} described the experimental setup followed to instantiate the generation framework within the context of pulsar observed by NenuFAR. The simulation parameters are specified, eight simulation scenarios are introduced and some implementation details regarding the network training are given. The experimental results are reported in Section \ref{Experimental results}, with respect to two different objectives, namely dynamic spectrum restoration and RFI detection. In particular, the performance of RFI-DRUNet in terms of RFI flagging is compared to those reached by state-of-the-art algorithms. Finally, in Section \ref{Estimaion of TOA}, an application of the proposed method to the estimation of pulsar time-of-arrival is presented. Section \ref{sec:conclusion} concludes the paper.  {For the sake of reproducibility and to promote open science, the data and the codes associated with this work are freely available online\footnote{\url{https://github.com/llxzhang/RFI-DRUnet}}.}

\section{Related works}
\label{Related Work}
In radio astronomy, numerous approaches have been proposed to mitigate the RFI during the post-correlation stage. These methods can be divided into two main categories, namely the parametric methods, and the data-driven methods.

Regarding the first type of approaches, \cite{fridman2001rfi} draw an overview of various types of RFI and RFI mitigation methods. Some RFI identification methods relying on statistical analysis tools have been proposed by \cite{fridman2008statistically}, \cite{bhat2005radio}, and \cite{winkel2007rfi}.  {In an earlier work, \cite{maslakovic1996excising} applied a thresholding technique after modeling the temporal waveform signal thanks to a discrete wavelet transform.  Besides, thresholding has also motivated the development of simple algorithms based on the general assumption that RFI are characterized by higher amplitudes than astronomical data. The popular cumulative sum method (CUSUM) initially proposed by \cite{page1954continuous} in the context of statistical process control was first applied by \cite{baan2004radio} to the RFI detection task.} Subsequently \cite{offringa2010post} and \cite{offringa2012morphological} have proposed several improvements referred to as VarThreshold, SumThreshold and AOFlagger. Notably, SumThreshold is a widely deployed algorithm for RFI removal in various current pipelines of radio telescopes because of its reliability and its efficiency \citep{offringa2010lofar,peck2013serpent,akeret2017hide}. {\cite{athreya2009new} has exploited the particular behavior of the fringe-stopped correlator output of an interferometer baseline in presence of RFI to remove spatially and temporally constant RFI sources. When dealing with multibeam receiver systems, \cite{kocz2010radio} have applied  spatial filtering to effectively identify and remove RFI from the temporal signals. The technique relies on a singular value decomposition (SVD) of the empirical covariance matrix computed from the Fourier representations of the input signals.  \cite{pen2009gmrt} have also attempted to identify RFI with the help of SVD, as well as \cite{zhao2013windsat} using a principal component analysis. More recently, \cite{finlay2023trajectory} have exploited the expected trajectories followed by RFI to tackle the problem of their removal jointly with the calibration task. Finally, it is worth noting that {Coastguard} \citep{lazarus2016prospects} and {Clfg} \citep{morello2019high} are two popular RFI mitigation algorithms specifically designed to remove RFI from pulsar data. They exploit the expected pulsar characteristics in combination with various common statistical tools.}

Conversely, data-driven approaches attempt to learn the main characteristics or features of RFI from existing data sets. Once adjusted, these models are deployed to identify the RFI. Such techniques rely on conventional machine learning tools such as K-nearest neighbors, Gaussian mixture models, and random forest, as employed by \cite{mosiane2017radio} and \cite{wolfaardt2016machine}. More recently, the last decade has been marked by the advent of deep neural networks, initially to perform vision-oriented tasks and then extended in various application domains. In this context, \cite{akeret2017radio} first designed a specific deep convolutional architecture, namely U-Net, to formulate  RFI identification as an image segmentation task. To improve the model capability, \cite{yang2020deep} introduce residual blocks and batch normalization, leading to the so-called RFI-Net specifically designed to identify RFI in data provided by the FAST radio telescope. \cite{yan2021radio} have investigated the relevance of \emph{atrous} convolution by proposing AC-UNet. R-Net proposed by \cite{vafaei2020deep} have demonstrated some robustness on simulated and real data by using transfer learning. {\cite{chang2023removing} propose DAARE, a stacked autoencoder model, to remove RFI from auroral kilometric radiation (AKR) spectrograms and to restore RFI-free astronomical data.} It is worth noting that the methods discussed above have been proposed in a supervised framework, i.e., they rely on available labeled data sets to train the designed networks. For instance \cite{Hamid2022PSRFINETRF} have used the predictions provided by the parametric methods Coastguard and Clfg as ground truth to train  PSRFINET for detection of the RFI in pulsar data. Other approaches have been developed to tackle the RFI mitigation problem in a semi- or non-supervised framework, i.e., when the data sets are not (or only partially) accompanied by labels.  For instance, \cite{ghanney2020radio} have compared the performance of the supervised method YOLO3 to those reached by an unsupervised method based on a convolutional autoencoder. \cite{mesarcik2020deep} have adopted a convolution variational autoencoder (VAE) and a naive support vector machine classifier to project high-dimensional time-frequency data into a low-dimensional prescriptive space. {\cite{mesarcik2022learning} have introduced an unsupervised method coined as nearest latent neighbors (NLN) which relies on a generative adversarial model to detect and identify RFI without direct observation of the interference. Motivated by a reduction of the computational cost, \cite{kerrigan2019optimizing} have proposed a network referred to as DFCN with comprehensible amplitude and phase information. Further, \cite{li2021detection} have proposed  RFI-GAN based on a generative adversarial network (GAN) and \cite{vos2019generative} have attempted to separate RFI signals from astronomical signals by using a source separation technique also based on a GAN architecture. \cite{wang2020radio} have implemented pseudoinverse learning autoencoders not only to remove RFI from pulsar data but also to restore the pulsar data possibly masked by RFI. Finally, \cite{saliwanchik2022self} have exploited the expected differences between RFI and astronomical signal with respect to their statistical representations and their resulting compressibility  to propose a self-learning network able to remove RFI.}


\section{A framework for generating RFI-corrupted dynamic spectra}
\label{simulation framework}

\subsection{Observation model for dynamic spectra}
\label{Observation model for dynamic spectra}
\begin{figure}[htbp]
    \centering
    \includegraphics[width=0.95\columnwidth]{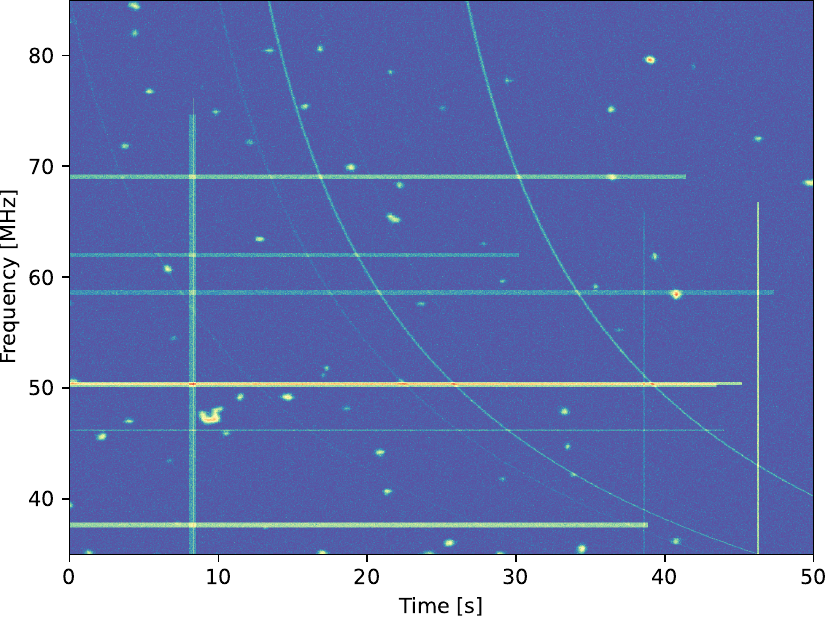}   
    \caption{An example of the dynamic spectrum generated according to the proposed model in Eq.~\eqref{eq:signal_model}.}
     \label{fig:tfplane}
\end{figure}

NenuFAR (New extension in Nancay upgrading LOFAR) is a new radio telescope built at the Nancay Radio Observatory, which is designed to observe the largely unexpected frequency window from $10$ to $85$MHz. NenuFAR currently operates according to several modes for the observation of pulsars, including folded mode, single-pulse mode, waveform mode, and dynamic spectra mode. In this work, we propose to identify and remove RFI from data acquired in the dynamic spectra mode, which basically consists of the magnitude of the discrete Fourier transform (DFT) of the raw data. Adopting a discretized, windowed analysis, dynamic spectra can be represented as a time-frequency plane, whose time and frequency resolutions are directly determined by the parameters of the DFT. The signal $S(\indFreq,\indTime)$ recorded in the $\indFreq$th frequency channel at the $\indTime$th temporal bin is assumed to be decomposed as 
\begin{equation} \label{eq:signal_model}
   S(\indFreq,\indTime) = P(\indFreq,\indTime)  + R(\indFreq,\indTime) + E(\indFreq,\indTime)  
\end{equation}
where $P(\indFreq,\indTime)$ is the pulsar signal, $R(\indFreq,\indTime)$ is a possible RFI component and $E(\indFreq,\indTime)$ stands for the system noise and any mismodeling, for $\indFreq \in \{ 1, \dots ,\NFreq \}$ and $\indTime \in \{ 1, \dots ,\NTime \}$. {Within the particular applicative context of the  work reported in this manuscript (i.e., NenuFAR-like tied-array observations), the model derived in Eq.~\eqref{eq:signal_model} is expected to be a reliable approximation of the dynamic spectra. For other particular contexts, this model could be enriched to account for various instrumental and acquisition parameters. For instance, the simulator specifically developed for the Hydrogen Epoch of Reionization Array (HERA) and considered by \cite{mesarcik2022learning} in the context of RFI mitigation offers the possibility of handling antenna cross-coupling \citep{herasim2024}.} An example of dynamic spectrum generated according to the model in Eq.~\eqref{eq:signal_model} is depicted in Fig.~\ref{fig:tfplane}, where the pulsar signature $P(\indFreq,\indTime)$ appears as exponentially decreasing curves and RFI $R(\indFreq,\indTime)$ take the form of vertical or horizontal lines and small clustered-dots. The remaining of this section will detail the procedures to generate each of these three components of the recorded signals. The notations used throughout this manuscript are summarized in Table \ref{tb:para_notation}.

\begin{table}[ht]\small
\centering
\begin{tabular}{|cl|}
\hline
\multicolumn{2}{|c|}{\textbf{Time-frequency representation}}                         \\ \hline
\multicolumn{1}{|l|}{$\NFreq$}         & Number of spectral bins (channels)            \\ \hline
\multicolumn{1}{|l|}{$n$}         & Index of the spectral bin (channel)            \\ \hline
\multicolumn{1}{|l|}{$\NTime$}         & Number of temporal bins                 \\ \hline
\multicolumn{1}{|l|}{$k$}         & Index of the temporal bin                \\ \hline
\multicolumn{1}{|l|}{$\delta t$}         & Temporal resolution                \\ \hline
\multicolumn{1}{|l|}{$\delta f$}         & Spectral resolution                \\ \hline
\multicolumn{1}{|l|}{$S(\indFreq,\indTime)$}      & Dynamic spectrum (full signal)                      \\ \hline
\multicolumn{1}{|l|}{$P(\indFreq,\indTime)$}      & Pulsar signal                          \\ \hline
\multicolumn{1}{|l|}{$R(\indFreq,\indTime)$}      & RFI signals                               \\ \hline
\multicolumn{1}{|l|}{$E(\indFreq,\indTime)$}      & System noise                           \\ \hline\hline
\multicolumn{2}{|c|}{\textbf{Pulsar modeling}}                                 \\ \hline
\multicolumn{1}{|l|}{$D$}           & Number of periods in the observation window  \\ \hline
\multicolumn{1}{|l|}{$d$}           & Index of the period in the observation window  \\ \hline
\multicolumn{1}{|l|}{$\rho$}   & Pulsar period                            \\ \hline
\multicolumn{1}{|l|}{$A(\indFreq,\indTime)$} & Integrated profile               \\ \hline
\multicolumn{1}{|l|}{$\gamma_{\mathrm{P}}(\cdot;\sigma^2)$} & $1$D Gaussian kernel \\ \hline
\multicolumn{1}{|l|}{$\textrm{SNR}_d$}  & Signal-to-noise ratio in the $d$th period   \\ \hline
\multicolumn{1}{|l|}{$\tau_n$} & Dispersion delay                          \\ \hline
\multicolumn{1}{|l|}{$\mathrm{DM}$}    &  Dispersion measure                       \\ \hline
\multicolumn{1}{|l|}{$L$}      & Number of Gaussian kernels          \\ \hline
\multicolumn{1}{|l|}{$a_\ell$}    & Amplitude of the $\ell$th Gaussian kernel                 \\ \hline
\multicolumn{1}{|l|}{$\sigma_\ell$}& Width of the $\ell$th Gaussian kernel           \\ \hline
\multicolumn{1}{|l|}{$\mu_\ell$}  &  Location of the $\ell$th Gaussian kernel              \\ \hline \hline
\multicolumn{2}{|c|}{\textbf{RFI modeling}}                                    \\ \hline
\multicolumn{1}{|l|}{$J$}      & Number of {RFI}                   \\ \hline
\multicolumn{1}{|l|}{$\textrm{SNR}_j$}  & Signal-to-noise ratio of the $j$th RFI      \\ \hline
\multicolumn{1}{|l|}{$M_j(\indFreq,\indTime)$} & Binary mask to locate the $j$th RFI \\ \hline
\multicolumn{1}{|l|}{$\gamma_{\mathrm{R}}(\cdot,\cdot;\sigma^2_{\mathrm{T}},\sigma^2_{\mathrm{F}})$} & Separable $2$D Gaussian kernel \\ \hline
\multicolumn{1}{|l|}{$\alpha$}      & Probability of occurrence of {nbct} and {bbt} {RFI}  \\ \hline
\multicolumn{1}{|l|}{$\beta$}      & Granularity parameter to generate {nbt} {RFI} \\ \hline
\end{tabular}
\caption{Notations used to describe the RFI-corrupted dynamic spectra generated by the simulation protocol.}
\label{tb:para_notation}
\end{table}

\subsection{Simulation of the pulsar signal}
\subsubsection{A template-based model}
\begin{figure}[htbp]
\centering
\includegraphics[width=0.95\columnwidth]{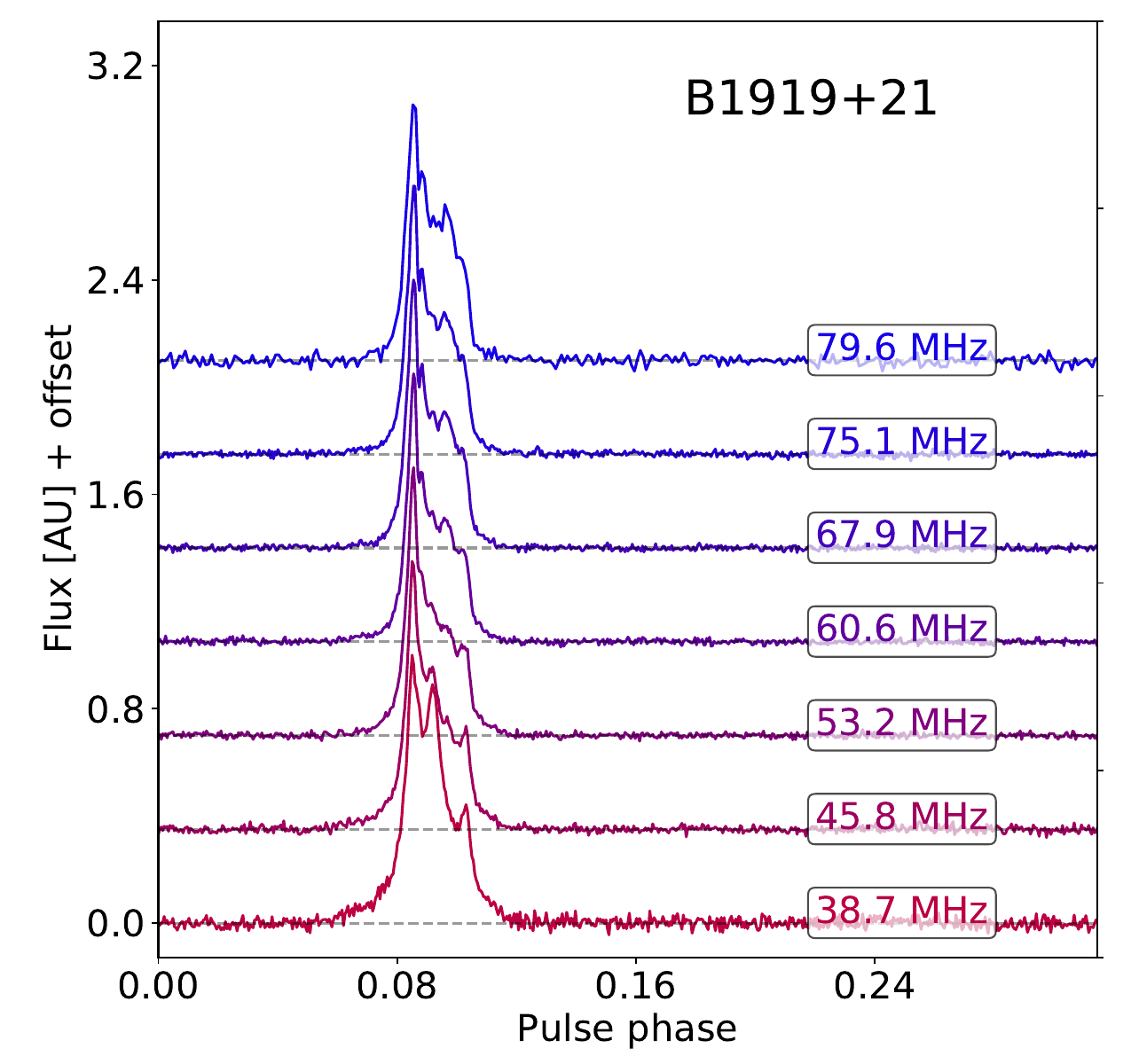}
\caption{Pulse profiles of the pulsar B1919+21 recorded at different frequencies by NenuFAR. The total integration time for the observation is $19.5$ minutes.  }
\label{fig:pulsar_profiles}
\end{figure}

The signal $P(\indFreq,\indTime)$ associated with the pulsar is mainly characterized by the so-called integrated pulse profile denoted $A(\indFreq,\indTime)$ hereafter. This profile can be considered as the "fingerprint" of a pulsar and is of primary interest to astronomers. Apart from possible variations in terms of energy, this signal generally shows fairly high stability in its shape along the observation at the same radio frequency but may exhibit small variations across the observation frequency \citep{Lorimer2004book}. To illustrate, Fig.~\ref{fig:pulsar_profiles} depicts profiles associated with the pulsar B1919+21 at different frequencies as observed by the NenuFAR telescope. It shows that the integrated pulse profile only slightly varies across  frequencies

Without loss of generality of the method developed throughout this manuscript, the pulsar profile is assumed to be fully described by a unique template that does not vary across frequencies. This template is chosen as a weighted linear combination of $L$ Gaussian-shaped components
\begin{equation}\label{eq:pulsar_profile}
    A(\indFreq,\indTime) =  \sum^{L}_{\ell=1} a_{\ell}  \gamma_{\mathrm{P}}(\indTime-\mu_\ell;\sigma_{\ell}^2)
\end{equation}
where $L$ is the number of Gaussian components composing the profile, $a_\ell$, $\mu_\ell$, and $\sigma_{\ell}^2$ stand for the amplitude, location and width of the $\ell$th component, respectively, and $\gamma_{\mathrm{P}}(t;\sigma^2) = \exp\left(-\frac{t^2}{2 \sigma^2}\right)$.  Observing  the pulsar over $D$ periods leads to a time-periodic pulsar signal written as
\begin{equation}\label{eq:pulsar_signal}
    P(\indFreq,\indTime) = \sum_{d=1}^{D} \mathrm{SNR}_d \times A(\indFreq, \indTime - \tau_n - d \rho )
\end{equation}
where $\mathrm{SNR}_d$ adjusts the signal-to-noise ratio in the $d$th period, $\rho$ is the pulsar period and $\tau_n$ is the frequency-varying delay resulting from the dispersion phenomenon. This quantity is discussed in the following paragraph.

\subsubsection{Dispersion measurement}
As radio pulses propagate through the interstellar medium and possibly the intergalactic medium, they are affected by a dispersive delay. The dynamic spectra mode operated by the radiotelescope can follow a so-called de-dispersion process to mitigate the impact of this delay through the frequency. However, for the sake of generality, the proposed simulated model for the pulsar signal is designed such that it accounts for this phenomenon. In this case, the dispersion measurement should be adjusted and, for frequency bin $n$, is given by 
\begin{equation}
    \tau_n = 4.15 \times \frac{\mathrm{DM}}{\indFreq^2\delta f^2\delta t}
\end{equation}
where $\delta f$ and $\delta t$ are the spectral and temporal resolutions of the dynamic spectrum. The dispersion measure (DM) is the electron column density through which the pulse has propagated.

\subsection{Simulation of {RFI}}\label{subsec:simulation_RFI}
Within the framework of acquisitions operated by NenuFAR, {RFI} can be roughly classified into three distinct types according to their respective shape along the frequency and the time domains: {\emph{i}) {narrow-band transient} ({nbt}) {RFI}, \emph{ii}) {narrow-band continuous-time} ({nbct}) {RFI} and \emph{iii}) {broad-band transient} ({bbt}) {RFI}.} To reflect this diversity in terms of RFI spatial and spectral patterns, the approach adopted in this work consists of decomposing the whole RFI signal $R(n,k)$ as the superimposition of $J$ individual RFI signatures. Each signature is described by a unique 2-dimensional Gaussian-shaped template $\gamma_R(t,f;\sigma^2_{\mathrm{T}}, \sigma^2_{\mathrm{F}}) \triangleq \exp\left(-\frac{t^2}{2\sigma^2_{\mathrm{T}}}\right)\exp\left(-\frac{f^2}{2\sigma^2_{\mathrm{F}}}\right)$ whose variances $\sigma^2_{\mathrm{T}}$ and $\sigma^2_{\mathrm{F}}$ adjust the temporal and the spectral spread of the pattern, respectively. {RFI} is assumed to be not affected by the dispersion effect, i.e., $\mathrm{DM}=0 \mathrm{pc\ cm}^{-3}$.  The whole RFI signal can be written as
\begin{equation}\label{eq:model_RFI}
    R(\indFreq,\indTime)= \sum_{j=1}^{J} \mathrm{SNR}_j \times \gamma_R(\indFreq-\indFreq_j,\indTime-\indTime_j;\sigma^2_{n_j},\sigma^2_{k_j})
\end{equation}
where $\mathrm{SNR}_j$ adjusts the power of the $j$th RFI, $\indFreq_j$ and $\indTime_j$ locate the spectral and temporal positions of the center of the $j$th RFI whose spectral and temporal spreads are driven by $\sigma^2_{n_j}$ and $\sigma^2_{k_j}$, respectively. To randomly locate the {RFI} over the dynamic spectra, the RFI model in Eq.~\eqref{eq:model_RFI} can be conveniently rewritten by explicitly introducing binary masks that are randomly drawn with prescribed statistical characteristics to mimic the diversity of the time-frequency shapes of the RFI. It yields
\begin{equation}
    R(\indFreq,\indTime)= \sum_{j=1}^{J} \mathrm{SNR}_j \times M_j(\indFreq,\indTime) \ast \gamma_R(\indFreq,\indTime;\sigma^2_{n_j},\sigma^2_{n_k})
\end{equation}
where $\ast$ stands for the 2-dimensional convolution operator and the binary mask
\begin{equation} \label{eq:RFI_masks}
M_j(\indFreq,\indTime) = \delta(\indFreq-\indFreq_j,\indTime-\indTime_j) = \begin{cases}
                1, & \text{if $\indFreq=k_j$ and $\indTime=n_j$}\\
                0, & \text{otherwise}
            \end{cases}
\end{equation}
takes the value $1$ in case of a RFI centered at the time instant $k_j$ in the frequency bin $n_j$, and $0$ otherwise.  Two different approaches have been followed to randomly generate these masks, each associated with particular types of {RFI}. These generation procedures are discussed in what follows.

\subsubsection{Spectrally and temporally extended {RFI}}\label{subsubsec:extended_RFIs}
{Narrow-band continuous-time} and {broad-band transient} {RFI} is two kinds of {RFI} that are ubiquitous in real observations. For instance, in observations made by NenuFAR, the 36-37MHz frequency band is often affected by such types of {interference}. Since these instances of RFI are generally mutually independent, it seems reasonable to assume that the entries of the corresponding masks $M_j(n,k)$ can be randomly generated according to a Bernoulli distribution with probability $\alpha$, i.e.,
\begin{equation*}
    \mathbb{P}[M_j(\indFreq,\indTime)=\epsilon] = \alpha^{\epsilon}(1-\alpha)^{1-\epsilon}
\end{equation*}
with $\epsilon\in \left\{0,1\right\}$. It is worth noting that  $\alpha = \mathbb{E}[M(\indFreq,\indTime)]$ is the probability of RFI occurrence in the dynamic spectra. Thus, this parameter directly adjusts the average number $\alpha N K$ of such spectrally and temporally extended ({nbct} and {bbt}) {RFI}. 

\subsubsection{{Narrow-band transient} {RFI}} 
To allow {nbt} {RFI} to affect the dynamic spectra in a clustered manner, the entries of the corresponding masks are not independently generated. Instead, they are assigned a Markov random field (MRF) which introduces structure correlation across the dynamic spectra. More precisely, this MRF is a multilevel logistic model (also known as an Ising model) and is defined as \citep{Li2009book}
\begin{multline}\label{eq:mask_MRF}
    \mathbb{P}\left[M_j(\indFreq,\indTime)=\epsilon \mid \mathcal{M}_j(n,k)\right] \\ \propto \alpha^{\epsilon}(1-\alpha)^{1-\epsilon} \exp\left[\beta \sum_{m \in \mathcal{M}_j(n,k)} \delta\left(m-\epsilon\right)\right]
\end{multline}
where $\epsilon\in \left\{0,1\right\}$ and
\begin{equation*}
\begin{split}
     \mathcal{M}_j(n,k) = \left\{{M}_j(n-1,k),{M}_j(n+1,k),\right.\\
     \left.{M}_j(n,k-1),{M}_j(n-1,k)\right\}
\end{split}    
\end{equation*}
denotes the set of neighbors of the mask entry $M_j(\indFreq,\indTime)$ in the time-frequency plane according to a 4-order neighboring structure. In Eq.~\eqref{eq:mask_MRF}, the so-called granularity parameter $\beta$ adjusts the correlation between neighboring entries in the mask. In particular, when $\beta=0$, no correlation is imposed and the model in Eq.~\eqref{eq:mask_MRF} reduces to the independent Bernoulli distributed model adopted for spectrally and temporally extended {RFI} (see Section \ref{subsubsec:extended_RFIs}). Simulating masks according to this MRF can be easily conducted thanks to Gibbs sampling \citep[Chap. 2]{Li2009book}.

\begin{figure*}[htbp]
\centering
\includegraphics[width=0.9\textwidth]{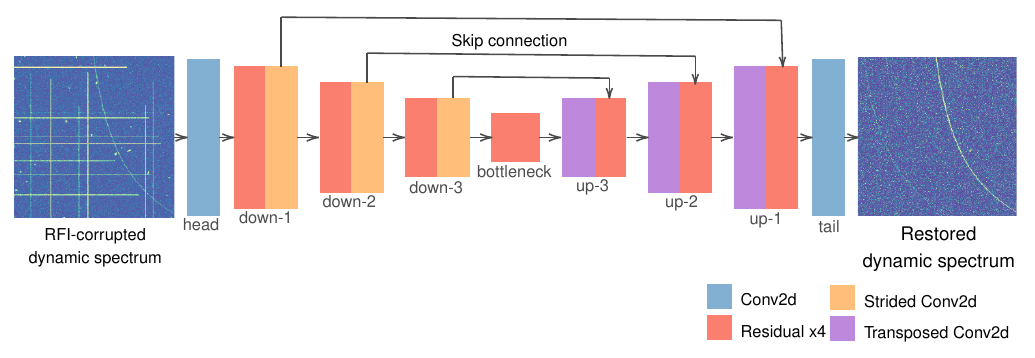}
\caption{Architecture of the proposed RFI-DRUnet network. It takes as inputs RFI-corrupted dynamic spectra and provides as output restored (i.e., RFI-free) dynamic spectra. Details about the layers are provided in Table \ref{tb:network_details}.}
\label{fig:network}
\end{figure*}

\subsection{Generation of simulated data sets}
\label{Generation of simulated data sets}
Thanks to the particular decomposed form of the proposed dynamic spectrum model in Eq.~\eqref{eq:signal_model}, the two main components associated with the pulsar signature and the RFI can be first generated independently and then combined to form the simulated signals. For instance, a given generated pulsar signal $P(\indFreq,\indTime)$ can be combined with several RFI signals $R(\indFreq,\indTime)$, or reciprocally, to produce an extended set of dynamic spectra $S(\indFreq,\indTime)$. Thus, in a nutshell, the proposed simulation framework consists in \emph{i}) generating a pulsar database $\mathcal{P}$ composed of pulsar signals, \emph{ii)} generating an RFI database $\mathcal{R}$ composed of RFI signals from three subsets,  $\mathcal{R}^{\mathrm{{nbt}}}$, $\mathcal{R}^{\mathrm{{nbct}}}$ and $\mathcal{R}^{\mathrm{{bbt}}}$, corresponding to three types of {RFI}, \emph{iii)} randomly selecting a pulsar signal from the database $\mathcal{P}$ and a RFI signal from the database $\mathcal{R}$ and \emph{iv)} combining them according to Eq.~\eqref{eq:signal_model} to build one RFI-corrupted dynamic spectra of data set $\mathcal{S}$. {The main benefit of this simulation strategy is to significantly reduce the amount of required computing. Indeed, let $|\mathcal{P}|$ and $|\mathcal{R}|$ denote the sizes of the pulsar and RFI databases, respectively. By coupling these two databases, the possible number of distinct generated dynamic spectra is 
$|\mathcal{S}| = |\mathcal{R}| \times |\mathcal{P}|$. In other words, only $|\mathcal{R}| + |\mathcal{P}|$
independent computations are necessary to generate $|\mathcal{S}|$ dynamic spectra, with $|\mathcal{R}| + |\mathcal{P}| \ll |\mathcal{S}|$.}

\section{Proposed RFI-DRUnet network for RFI mitigation}
\label{Deep denoising network}
\subsection{Formulating RFI mitigation as image restoration}
\label{Denoising method}
As stated in Section \ref{introduction}, most of the methods proposed in the literature have formulated the RFI mitigation problem as a detection or segmentation/classification task. In other words, they only aim at localizing the time-frequency bins of the dynamic spectra possibly affected by  {RFI}, distinguishing RFI-corrupted bins from RFI-free bins. Conversely, we propose to go beyond this crude RFI detection by formulating the problem of RFI mitigation as a restoration task. Our main rationale is that recent advances in machine learning offer the possibility of recovering clear (i.e., RFI-free) dynamic spectra from the RFI-corrupted measurements directly. To do so, we interpret the RFI mitigation objective as image denoising enounced in the time-frequency plane. Image denoising, a particular instance of image restoration, consists of recovering a clean image ${X}$ from the degraded image ${Y}$ corrupted by a specific (measurement) noise. When this noise is assumed to be additive and denoted $\tilde{{E}}$, the clean and corrupted images are related through the observation model
\begin{equation}\label{eq:denoising_general}
    {Y} =  {X}  + \tilde{E}.
\end{equation}
It is worth noting that this generic formulation of the denoising task under the model in Eq.~\eqref{eq:denoising_general} perfectly matches the objective of recovering RFI-free spectra from dynamic spectra under the model in Eq.~\eqref{eq:signal_model}. To draw this connection, the observed image $Y$ (resp. clean image $X$) in Eq.~\eqref{eq:denoising_general} can be associated with the measured dynamic spectra $S$ (resp. RFI-free spectra $P+E$) in Eq.~\eqref{eq:signal_model} while the noise term $\tilde{E}$ writes $\tilde{E}=R$ in the context of RFI mitigation. In other words, RFI mitigation consists of restoring the dynamic spectrum corrupted by a particularly type of structured noise. As a benefit of translating RFI mitigation into image denoising, one can easily capitalize on recently proposed powerful deep convolutional neural networks such as those proposed by \cite{zhang2017beyond}, \cite{jiang2018rednet} and \cite{zhang2020residual}. The architecture of the proposed network, hereafter referred to as RFI-DRUnet, is detailed in what follows.

\subsection{RFI-DRUnet architecture}
Image denoising is an archetypal image-to-image translation task, which can be efficiently addressed using deep networks with encoder-decoder architectures since they consistently deliver exceptional results. The encoder of the network is able to extract input image features at various levels while reducing the data size, and the decoder reconstructs data with the aid of features at different levels provided by so-called skip connections. {In this work, we have customized the popular network referred to as DRUNet \citep{zhang2021plug} to be in agreement with the targeted RFI mitigation task.} DRUNet is a deep convolutional network that follows an encoder-decoder architecture and utilizes a residual module that notably enhances the network capacity of feature extraction. {Conventional DRUNet implementations are able to handle different noise levels in the data. Indeed, during the training stage, the noisy image and a corresponding noise map are jointly provided as the inputs of the network. For the RFI mitigation task considered in this work, since {interference} to be removed from the dynamic spectra can be of any intensity level, such a flexibility is not required. Thus the RFI-DRUnet architecture is designed such that only RFI-corrupted dynamic spectra are provided as inputs during the training stage. In other words, no noise map is jointly provided as input to the proposed network during training.}

More specifically, the architecture of the proposed RFI-DRUNet is sketched in Fig.~\ref{fig:network}. Its backbone consists of three parts, namely the encoder, the decoder, and the middle connection layer. The encoder and decoder networks are mirrored, i.e.,  each module in the encoder has an associated counterpart in the decoder and is connected by skip-connection. After passing a 64-channel convolutional (head) layer, the encoder comprises three modules, each consisting of 4 residual blocks and a stridden convolution (stride $2 \times 2$) as the downsampling layer. The middle connection layer is stacked by 4 residual blocks, which are followed by a decoder structurally symmetrical to the encoder whose modules are sequentially composed of a transposed convolution (stride $2 \times 2$, padding $2 \times 2$) as the upsampling layer and 4 residual blocks. After the decoder, the network ends with a convolutional (tail) layer with one channel.

All residual blocks in the network are composed of 2 residual layers connected by a ReLU activation function. Furthermore, as in the conventional implementation of DRUNet, there is no activation function except in the residual blocks. From an overall perspective, the number of channels starts from 1 and increases to 64 after the first convolutional layer, then it is doubled with each downsampling layer and halved for each up-sampling layer until 64 then becomes 1 through the last convolutional layer. The details of each layer of the network are summarized in Table \ref{tb:network_details}.

\begin{table*}\small
\centering
\begin{tabular}{c|c|c|c|c|c|c|c}
 \hline
{\textbf{Layer}} &
  \textbf{Block} &
  \textbf{Operation} &
  \textbf{Kernel size} &
  \textbf{Stride} &
  \textbf{Padding} &
  \textbf{Input size} &
  \textbf{Output Size} \\ \hline
{head} &
 {Conv2d} &
  {Conv2d} &
  {$3 \times 3 \times 64$} &
  {$(1,1)$} &
  {$(1,1)$} &
  {$(64,64,1)$} &
  {$(64,64,64)$} \\ \hline
\multirow{ 2}{*}{down-1} &
  {Residual $\times 4$} &
  {(Conv2d+ReLu+Conv2d) $\times 4$} &
  {$3 \times 3 \times 64$} &
  {$(1,1)$} &
  {$(1,1)$} &
  {$(64,64,64)$} &
  \multicolumn{1}{|c}{$(64,64,64)$} \\
  &
  {Strided Conv2d} &
  {Conv2d} &
  {$2 \times 2 \times 128$} &
  {$(2,2)$} &
  {$(1,1)$} &
  {$(64,64,64)$} &
  \multicolumn{1}{|c}{$(32,32,128)$} \\ \hline
\multirow{ 2}{*}{down-2} &
  {Residual $\times 4$} &
  {(Conv2d+ReLu+Conv2d) $\times 4$} &
  {$3 \times 3 \times 128$} &
  {$(1,1)$} &
  {$(1,1)$} &
  {$(32,32,128)$} &
  \multicolumn{1}{|c}{$(32,32,128)$} \\
 &
  {Strided Conv2d} &
  {Conv2d} &
  {$2 \times 2 \times 256$} &
  {$(2,2)$} &
  {$(1,1)$} &
  {$(32,32,128)$} &
  \multicolumn{1}{|c}{$(16,16,256)$} \\ \hline
\multirow{ 2}{*}{down-3} &
  {Residual $\times 4$} &
  {(Conv2d+ReLu+Conv2d) $\times 4$} &
  {$3 \times 3 \times 256$} &
  {$(1,1)$} &
  {$(1,1)$} &
  {$(16,16,256)$} &
  \multicolumn{1}{|c}{$(16,16,256)$} \\
 &
  {Strided Conv2d} &
  {Conv2d} &
  {$2 \times 2 \times 512$} &
  {$(2,2)$} &
  {$(1,1)$} &
  {$(16,16,256)$} &
  \multicolumn{1}{|c}{$(8,8,512)$} \\ \hline
{{bottleneck}} &
  {Residual $\times 4$} &
  {(Conv2d+ReLu+Conv2d) $\times 4$} &
  {$3 \times 3 \times 512$} &
  {$(1,1)$} &
  {$(1,1)$} &
  {$(8,8,512)$} &
  \multicolumn{1}{|c}{$(8,8,512)$} \\ \hline
\multirow{ 2}{*}{up-3} &
  {Transposed Conv2d} &
  {Conv2d} &
  {$3 \times 3 \times 256$} &
  {$(2,2)$} &
  {$(2,2)$} &
  {$(8,8,512)$} &
  \multicolumn{1}{|c}{$(16,16,256)$} \\
 &
  {Residual $\times 4$} &
  {(Conv2d+ReLu+Conv2d) $\times 4$} &
  {$3 \times 3 \times 256$} &
  {$(1,1)$} &
  {$(1,1)$} &
  {$(16,16,256)$} &
  \multicolumn{1}{|c}{$(16,16,256)$} \\ \hline
\multirow{ 2}{*}{up-2} &
  {Transposed Conv2d} &
  {Conv2d} &
  {$3 \times 3 \times 128$} &
  {$(2,2)$} &
  {$(2,2)$} &
  {$(16,16,256)$} &
  \multicolumn{1}{|c}{$(32,32,128)$} \\
 &
  {Residual $\times 4$} &
  {(Conv2d+ReLu+Conv2d) $\times 4$} &
  {$3 \times 3 \times 128$} &
  {$(1,1)$} &
  {$(1,1)$} &
  {$(32,32,128)$} &
  \multicolumn{1}{|c}{$(32,32,128)$} \\ \hline
\multirow{ 2}{*}{up-1} &
  {Transposed Conv2d} &
  {Conv2d} &
  {$3 \times 3 \times 64$} &
  {$(2,2)$} &
  {$(2,2)$} &
  {$(32,32,128)$} &
  \multicolumn{1}{|c}{$(64,64,64)$} \\
 &
  {Residual $\times 4$} &
  {(Conv2d+ReLu+Conv2d) $\times 4$} &
  {$3 \times 3 \times 64$} &
  {$(1,1)$} &
  {$(1,1)$} &
  {$(64,64,64)$} &
  \multicolumn{1}{|c}{$(64,64,64)$}\\ \hline
{tail} &
  Conv &
  Conv &
  $3 \times 3 \times 1$ &
  $(1,1)$ &
  $(1,1)$ &
  $(64,64,1)$ &
  \multicolumn{1}{|c}{$(64,64,1)$} \\ \hline
\end{tabular}
\caption{Details of the layers of the proposed RFI-DRUnet network. The size of input data and output data are $(H \times W \times C)$, where $H$, $W$ and $C$ stand for height, width and the number of channels, respectively. Sizes of the input can vary depending on the needs; an input data size of $64 \times 64 $ is chosen here as an example.}\label{tb:network_details}
\end{table*}

\section{Experimental framework} 
\label{Experiments framework}
{Extensive experiments have been conducted to assess the performance of the proposed RFI-DRUNet method. More precisely, the network designed in Section \ref{Deep denoising network}  will be trained on several synthetic data sets generated following the  simulation framework outlined in Section \ref{simulation framework}.} The parameters of the model are specified in Section \ref{subsec:Simulation parameters} and the simulation scenarios as well as associated data sets are described in Section \ref{Synthetic Datasets}. Finally, implementation details are provided in Section \ref{Implementation}.

\subsection{Simulation parameters}\label{subsec:Simulation parameters}

\noindent \emph{Parameters of the time-frequency representation --} The parameters of the dynamic spectra are chosen to match the main characteristics of the signals observed by NenuFAR. The dynamic spectra is assumed to be characterized by $N=1024$ frequency channels ranging from $35$MHz and $85$MHz with a spectral resolution of $\delta f=48.828$kHz and $K=1024$ temporal bins of resolution of $\delta t = 0.05$s.\\

\noindent \emph{Parameters of the pulsars --}  The signal of the pulsar $P(\indFreq,\indTime)$ is fully described in Eq.~\eqref{eq:pulsar_signal} by its integrated profile $A(\indFreq,\indTime)$, defined in Eq.~\eqref{eq:pulsar_profile} by the template $\gamma_{\mathrm{P}}(\cdot,\cdot)$, and its periodization over the observation time. The values of the pulsar period $\rho$ and the dispersion factor $\mathrm{DM}$  are uniformly and randomly drawn over specific ranges chosen to mimic realistic signals. The power levels $\mathrm{SNR}_d$ ($d=1,\ldots,D$) of the pulsar defined on the $D$ periods are selected according to a log-uniform rule. The number $L$ of Gaussian defining the template is at most equal to $2$, localization $\mu_\ell$ and width $\sigma^2_\ell$ of each Gaussian shape are uniformly drawn over pre-defined sets. The admissible ranges of the parameters are reported in Table \ref{tab:pulsar_parameters}.\\

\begin{table}[H]\small
 \renewcommand{\arraystretch}{1.1}
    \centering
    \begin{tabular}{lcc}
        \hline
       \textbf{Parameter}  & \textbf{Notation} & \textbf{Value range}\\
       \hline
       Period  & $\rho$ & $\left(20 , 40\right)$ [bins] \\
       Dispersion  & DM & $(10, 40)$ [pc cm$^{-3}$]\\
       Number of components & $L$ & $\{1,2\}$ \\
       Amplitude  & $a_\ell$ & $(0.2 , 1)$ \\
       Localization  & $\mu_\ell$ & $(0 , \rho)$ \\
       Width  & $\sigma_\ell^2$ & $(0.01 , 0.04)$\\
       Power  & $\mathrm{SNR}_d$ & $(0.01, 20)$ \\   
       \hline
    \end{tabular}
    \caption{Parameters of the simulation associated with the pulsar signal.}
    \label{tab:pulsar_parameters}
\end{table}

\noindent \emph{Parameters of the {RFI} --} The three types of {RFI}, namely {nbt}, {bbt} and {nbct}, are generated following the same simulation protocol. It consists in drawing binary masks according to a Bernoulli distribution ({bbt} and {nbct} {RFI}) or a Markov random field ({nbt} {RFI}). These binary masks are then convolved with a 2-dimensional separable Gaussian kernel whose spectral $\sigma^2_{\mathrm{F}}$ and temporal and $\sigma^2_{\mathrm{T}}$ spreading are randomly drawn over predefined sets chosen according to the type of generated {RFI}. The admissible ranges of the parameters involved in the Gaussian kernel as well as the parameters adjusting the statistical properties of the binary masks are reported in Table \ref{tab:RFI_parameters} for the three types of {RFI}.\\

\begin{table}[H]\small
\renewcommand{\arraystretch}{1.1}
\setlength{\tabcolsep}{4pt}
   \centering
\begin{tabular}{c c c c c c} 
\hline
\textbf{Type of RFI}  & $\mathrm{SNR}_j$  & $\sigma^2_{\mathrm{F}}$ & $\sigma^2_{\mathrm{T}}$ & $\alpha$  & $\beta$\\ \hline
{bbt}       & $(1,10)$       & $(600,1024)$        & $(1,10)$      & $(0,0.01)$ &  N/A\\
{nbct}       & $(1,10)$       & $(1,10)$            & $(600,1024)$  & $(0,0.01)$ &  N/A\\ 
{nbt}       & $(0,1)$        & $(1,11)$            & $(1,11)$      & $0.8$ &   $40$\\ \hline
\end{tabular}
    \caption{Parameters of the simulation associated with the RFI signal.}
    \label{tab:RFI_parameters}
\end{table}

\subsection{Simulation scenarios} 
\label{Synthetic Datasets}
As described in Section \ref{Generation of simulated data sets}, the generated dynamic spectra are combinations of pulsar signals from data set $\mathcal{P}$ and multiple RFI signals from data set $\mathcal{R}$. This simulation protocol is instantiated to produce three distinct data sets, namely the training set, the validation and the testing set. The training and validation sets share the same pulsar and  RFI databases, but differ by the size of $\mathcal{S}$. The testing set is built from different pulsar and RFI sets. The sizes of these sets are reported in Table \ref{tab:dataset_size}, where the RFI set $\mathcal{R}$ contains three subsets matching three types of RFI, each of the same size.

\begin{table}[ht!]\small
 \renewcommand{\arraystretch}{1.1}
 \setlength{\tabcolsep}{4pt}
    \centering
    \begin{tabular}{lccc}
      \hline
         & \textbf{Size of} $\mathcal{P}$  &  \textbf{Size of} $\mathcal{R}$ & \textbf{Size of} $\mathcal{S}$\\
      \hline
      Training set     & $20$   & $300$ & $1800$\\
      Validation set   & $20$   & $300$                             & $200$\\
      Testing set      & $10$   & $60$  & $200$\\
      \hline
    \end{tabular}
    \caption{Size of the generated training, validation and testing sets.}
    \label{tab:dataset_size}
\end{table}

\begin{figure}[ht!]
\centering
\includegraphics[width=0.95\columnwidth]{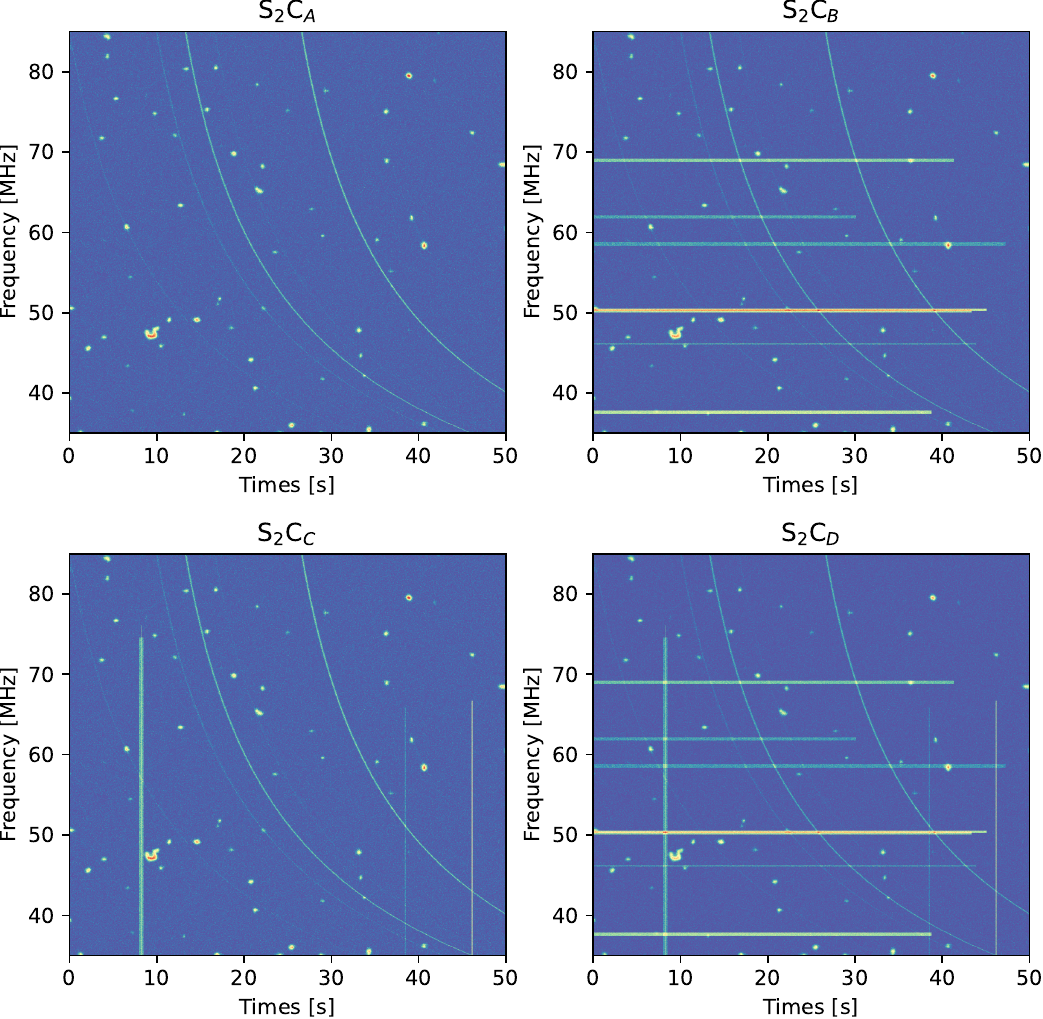}
\caption{Examples of the dynamic spectrum generated according to the proposed protocol for the 4 cases of $\mathsf{S}_2$}
\label{fig:data_4cases}
\end{figure}

According to this generation protocol, two experimental scenarios are considered to assess the robustness of the proposed method with respect to system noise. More precisely, so-called Scenario 1 (shortened as $\mathsf{S}_1$ hereafter) considers dynamic spectra composed of pulsar signal and {RFI}, i.e., free of system noise, $E=0$ in Eq.~\eqref{eq:signal_model}. {Conversely, Scenario 2 (shortened as $\mathsf{S}_2$ hereafter) considers dynamic spectra generated according to the full model in Eq.~\eqref{eq:signal_model}, i.e., composed of a pulsar signal, RFI and a system noise modeled as an additive white Gaussian variable with variance  $\sigma_{\mathrm{E}}^2 = 1$, corresponding to an average value of signal-to-noise ratio (SNR) of $-5.6$dB over the test set. It is worth noting that Scenario 1 has been considered in addition to Scenario 2 to assess the impact of the system noise on the performance of the proposed restoration method. This scenario will be also of interest to evaluate the best restoration performance that could be reached by an oracle RFI detector (see Section \ref{Denoising results}).}  Moreover, to further investigate the model capacity to manage various types of {RFI}, for each scenario, we consider 4 cases that differ by the composition of the RFI set $\mathcal{R}$. These cases, denoted as $\mathsf{C}_{\mathrm{A}}$ to $\mathsf{C}_{\mathrm{D}}$, are defined as follows
\begin{itemize}
    \item $\mathsf{C}_{\mathrm{A}}$: {nbt} {RFI} (pulse-like {RFI})
    \item $\mathsf{C}_{\mathrm{B}}$: {nbt} {RFI} + {nbct} {RFI} (narrow-band {RFI})
    \item $\mathsf{C}_{\mathrm{C}}$: {nbt} {RFI} + {bbt} {RFI} ({transient} {RFI})
    \item $\mathsf{C}_{\mathrm{D}}$: {nbt} {RFI} + {nbct} {RFI} + {bbt} {RFI} (all {RFI} types)
\end{itemize}
To summarize, a total of 8 simulation scenarios are considered, denoted as $\mathsf{S}_{{\small{\square}}}\mathsf{C}_{\triangle}$, with ${\small{\square}} \in \left\{1,2\right\}$ and $\triangle \in \left\{\mathrm{A}, \mathrm{B}, \mathrm{C}, \mathrm{D}, \right\}$, depending on the presence/absence of system noise and depending on the type of {RFI} corrupting the pulsar signal. To illustrate, Fig.~\ref{fig:data_4cases} displays one generated dynamic spectrum for each case in scenario $\mathsf{S}_2$.

\subsection{Implementation and training details} 
\label{Implementation}
{A distinct RFI-DRUNet model has been trained on each data set associated with the 8 simulation scenarios, leading to 8 instances of the proposed network.} For all models, the training parameters are the same. During training, the generated dynamic spectrum from the data set $\mathcal{S}$ has been randomly cropped to a patch of size $64 \times 64$. Data augmentation is then applied, including flipping and rotating. The $\ell_1$-norm  has been used to define the loss function. The Adam optimizer with a mini-batch size of 64 has been employed with a learning rate starting from $10^{-4}$ and halved after every 100k iterations until it reaches the value $5 \times 10^{-7}$. The training takes about $40$ hours with 10000 epochs for an implementation in Pytorch and equipped with an Nvidia RTX 3080 GPU. {This choice in terms of batch size and number of epochs has been driven by technical constraints imposed by the available computing resources. A similar strategy has been adopted by \cite{zhang2017beyond}.}

\section{Experimental results}
\label{Experimental results}
The experiments conducted following the framework described in the previous section mainly follow two objectives. Firstly, they have been designed to demonstrate the feasibility of restoring RFI-corrupted dynamic spectra. Appropriate denoising-inspired quantitative performance measures will thus be detailed in Section \ref{Evaluation metric}. {Secondly, we will show that the proposed restoration method can be easily simplified to perform only an RFI detection task.} It consists of converting the outputs of the proposed RFI-DRUNet method into binary masks locating the RFI-corrupted time-frequency bins. This allows the proposed approach to be compared to state-of-the-art methods such as U-Net \citep{akeret2017radio} and RFI-Net \citep{zhang2020residual} which have formulated RFI mitigation as a segmentation/classification task. When interpreted in this way, the methods can be compared with respect to detection-oriented figures-of-merit also defined in Section \ref{Evaluation metric}. The quantitative results are then reported and discussed in Sections \ref{Denoising results} and \ref{Comparison with classification methods} with respect to the two aforementioned tasks, namely restoration and detection. An illustration on a real observation is presented in Section \ref{Illustration on a real data}. Finally, in light of these {results}, Section \ref{Discussion} discusses the relevance of the restoration paradigm with respect to the conventional RFI mitigation approaches, namely detection/flagging. It also {mitigates} these encouraging results by highlighting some cases of recovery failures experienced by RFI-DRUNet.

\subsection{Quantitative figures-of-merit} 
\label{Evaluation metric}
As suggested above, the performance of the proposed method will be evaluated with respect to two main objectives, namely dynamic spectrum restoration and RFI detection (also referred to as RFI flagging in the literature). Regarding the restoration task, a conventional figure-of-merit encountered in image processing is the peak signal-to-noise ratio (PSNR) defined as
\begin{equation}
    \mathrm{PSNR} = 10\log_{10}\frac{\max X^2}{\mathrm{MSE}(X,\hat{X})}
\end{equation}
where $\max X^2$ is the (squared) maximum value of the RFI-free signal, and $\mathrm{MSE}(X,\hat{X})$ quantifies the mean square error between the ground truth RFI-free signal $X=S-R$ and its restored counterpart estimated by the algorithm $\hat{X}$. In particular, in the noise free case ($E=0$, Scenario 1), the observation model in Eq. \eqref{eq:signal_model} simplifies to $S=P+R$ and this metric boils down to measuring the quality of the restored pulsar signal, i.e., $\hat{X} = \hat{P}$.

Regarding the detection task, since it can be formulated as a binary classification problem, the performance of the mitigation method can be evaluated by resorting to conventional classification-oriented figures of merit. It requires to first computing the so-called confusion matrix which summarizes the numbers of correct and bad classifications with respect to the presence of a given target. For the RFI mitigation task of interest, we define the presence of an RFI as a positive instance and, conversely, an RFI-free signal as a negative instance. The confusion matrix reports the estimated probabilities of positive and negative samples being classified correctly or incorrectly, which leads to the definition of four indicators named \emph{true positive} (TP), \emph{true negative} (TN), \emph{false positive} (FP) and \emph{false negative} (FN). Once the confusion matrix has been computed, standard metrics can be derived, which include precision, recall, and F1 score. 
Precision measures the proportion of predicted positives that are true positives
\begin{align*}
    \mathrm{prec} = \frac{\mathrm{TP}}{\mathrm{TP}+\mathrm{FP}}.
\end{align*}
Recall is the percentage of correctly identified {RFI}   
\begin{align*}
    \mathrm{rec} = \frac{\mathrm{TP}}{\mathrm{TP}+\mathrm{FN}}.
\end{align*}
The F1 score, which strikes a balance between precision and recall, is calculated as the harmonic mean. This metric is especially valuable when analyzing data sets with unbalanced classes, i.e., when the number of samples in a given class is significantly larger than the number of samples in the other class, which is expected to be the case for moderately corrupted dynamic spectra. It is defined as
\begin{align*}
    \mathrm{F}1 = \frac{2 \times \mathrm{prec} \times \mathrm{rec}}{\mathrm{prec} + \mathrm{rec}}. 
\end{align*}
{Following the evaluation protocol also adopted by \cite{mesarcik2022learning}, the performance has been also evaluated in term of the area under the precision-recall curve (AUPRC), which quantifies the overall discriminatory ability of the compared models. This metric is also particularly valuable in scenarios with imbalanced datasets, where the number of negative instances significantly outweighs the positive ones. AUPRC emphasises the ability of the model in correctly identifying positive examples while maintaining accuracy.  Finally, the evaluation of the model performance has been conducted in light of the receiver operating characteristics (ROC), which provides a comprehensive description of the performance of any detector faced to a binary hypothesis testing. ROC curves plot the true positive rate (TPR or probability of detection) as a function of the false positive rate (FPR or probability of false alarm). The area under the ROC (AUROC) sometimes referred to as the c-statistic (Hastie et al., 2009) has been considered as a figure-of-merit.}


\subsection{Restoration results}
\label{Denoising results}
As described in Section \ref{Synthetic Datasets}, 8 simulation scenarios have been considered depending on the presence or absence of system noise and depending on the type of {RFI} corrupting the dynamic spectra. The proposed RFI-DRUNet model has been trained separately on 8 data sets associated with each of these scenarios. For each scenario $\mathsf{S}_1$ or $\mathsf{S}_2$, the restoration performances of the RFI-DRUNet trained for a given case ($\mathsf{C}_{\mathrm{A}}$ to $\mathsf{C}_{\mathrm{D}}$) are evaluated not only on a testing data set generated according to the same case but also on data sets corresponding to the other cases. This will help to understand the impact of the type of {RFI} on the model performance and its possible limitations. {With a slight abuse of notations, the models will be denoted as $\mathsf{S}_{{\small{\square}}}\mathsf{C}_{\triangle}$ by shortening the denomination of the data set they have been trained on, where the two indices ${\small{\square}} \in \left\{1,2\right\}$ and $\triangle \in \left\{\mathrm{A}, \mathrm{B}, \mathrm{C}, \mathrm{D}, \right\}$ refer to the considered scenario and case, respectively.}\\

\noindent {\emph{Noise-free data set --} Table \ref{tb:psnr_s1} reports the results of the restoration in terms of average PSNR and standard deviations computed over the test set for the 4 models trained for the scenario $\mathsf{S}_1$. To appreciate the performance gain in term of restoration, this table also reports the PSNR computed from the data itself (first row). This is the cheapest strategy that would consist in not performing any restoration of the data. It also provides a quantitative proxy of the difficulty of the RFI-mitigation task (the lower SNR, the more difficult task). Finally, the second row of the table reports the PSNR computed from the noise-free pulsar signal recovered by an oracle detector that would be able to perfectly identify the RFI and would replace the corrupted time-frequency bins by zeros.} These results show that the model $\mathsf{S}_{{{1}}}\mathsf{C}_{\mathrm{A}}$ performs quite differently, since it provides very good results only when tested on the data set $\mathsf{S}_{{{1}}}\mathsf{C}_{\mathrm{A}}$. This can be explained by the limited variety of {RFI}  (only of type {nbt}) in the training set. Other types of {RFI} are hardly identified and corrected by the algorithm. The models $\mathsf{S}_{{{1}}}\mathsf{C}_{\mathrm{B}}$ to $\mathsf{S}_{{{1}}}\mathsf{C}_{\mathrm{D}}$  perform well not only on the test sets corresponding to their training set but also show comparable restoration ability when handling the other cases. These three later cases contain pulse-like (i.e., {nbt}) {RFI} but differ substantially since they contain either {nbct} {RFI} or {bbt} {RFI}, or both. These two types of {RFI} share some shape similarities but differ in the direction of spreading. The data augmentation used during the training phase, which consists of rotations and flips, can explain this robustness to handle both {nbct} and {bbt} {RFI} when only one of them is present in the training set. \\

\begin{table}[ht!]\small
\setlength{\tabcolsep}{6pt}
\centering
\begin{tabular}{lc|cccc}
\hline
\multicolumn{2}{c|}{\diagbox{\textbf{Model}}{\textbf{Data set}}} & $\mathsf{S}_1\mathsf{C}_{\mathrm{A}}$ & $\mathsf{S}_1\mathsf{C}_{\mathrm{B}}$ & $\mathsf{S}_1\mathsf{C}_{\mathrm{C}}$ & $\mathsf{S}_1\mathsf{C}_{\mathrm{D}}$ \\ 
\hline
\multicolumn{2}{c|}{\multirow{2}{*}{Data}}   & $43.17$                       & $35.02$                 & $35.87$                      & $31.92$                  \\
&                                                          & \scriptsize $\pm 0.93$     & \scriptsize $\pm 3.14$    & \scriptsize $\pm 2.99$    & \scriptsize $\pm 1.33$\\
\hline
\multicolumn{2}{c|}{\multirow{2}{*}{Oracle}}   & $59.27$                       & $54.43$                 & $54.24$                      & $51.94$                  \\
&                                                          & \scriptsize $\pm 6.37$     & \scriptsize $\pm 6.11$    & \scriptsize $\pm 5.61$    & \scriptsize $\pm 5.51$\\
\hline 
\multicolumn{1}{c|}{\multirow{8}{*}{\rotatebox{90}{RFI-DRUNet} }}   &\multirow{2}{*}{$\mathsf{S}_1\mathsf{C}_{\mathrm{A}}$}  & 70.58         & 56.59        & 59.75       & 53.82 \\
\multicolumn{1}{c|}{} &                                                             & \scriptsize $\pm$10.86    & \scriptsize $\pm$10.04   & \scriptsize $\pm$8.96   & \scriptsize $\pm$ 8.10\\
\cline{2-6}
\multicolumn{1}{c|}{} &\multirow{2}{*}{$\mathsf{S}_1\mathsf{C}_{\mathrm{B}}$}  & 73.33         & 72.38        & 71.74       & 70.80 \\
\multicolumn{1}{c|}{} &                                                             & \scriptsize $\pm$8.70     & \scriptsize $\pm$8.24    & \scriptsize $\pm$8.30   & \scriptsize $\pm$ 7.80\\
\cline{2-6}                                                        
\multicolumn{1}{c|}{} &\multirow{2}{*}{$\mathsf{S}_1\mathsf{C}_{\mathrm{C}}$}  & 72.65         & 70.94        & 71.45       & 69.85 \\
\multicolumn{1}{c|}{} &                                                             & \scriptsize $\pm$9.00     & \scriptsize $\pm$8.32    & \scriptsize $\pm$8.65   & \scriptsize $\pm$ 7.93\\
\cline{2-6}                                                       
\multicolumn{1}{c|}{} &\multirow{2}{*}{$\mathsf{S}_1\mathsf{C}_{\mathrm{D}}$}  & 72.09         & 71.44        & 71.31       & 70.72 \\ 
\multicolumn{1}{c|}{} &                                                            & \scriptsize $\pm$9.23     & \scriptsize $\pm$8.87    & \scriptsize $\pm$8.92   & \scriptsize $\pm$ 8.61\\        
\hline                           
\end{tabular}
\caption{{Scenario 1: restoration performance in terms of average PSNR and standard deviations computed over the test data sets.}}\label{tb:psnr_s1}
\end{table}

\noindent \emph{Noisy data sets --} Similar findings can be drawn when considering the 4 models trained and tested on noisy data sets (scenario $\mathsf{S}_2$),  as shown in Table \ref{tb:psnr_s2}. {Note that, in this case, the time-frequency bins identified by the oracle have been replaced by random values drawn according to the noise statistical model, i.e., $\mathcal{N}(0,\sigma^2_{\mathrm{E}})$.} {The overall values of PSNR are significantly lower than obtained with scenarios $\mathsf{S}_1$, owing to the presence of system noise. However, the PSNR values are sufficiently high to guarantee correctly restored dynamic spectra, in regard to standard restoration measures encountered in the image processing literature but also to the performance reached by the oracle detector.}\\

\begin{table}[ht!]\small
\setlength{\tabcolsep}{6pt}
\centering
\begin{tabular}{lc|cccc}
\hline
\multicolumn{2}{c|}{\diagbox{\textbf{Model}}{\textbf{Data set}}} & $\mathsf{S}_2\mathsf{C}_{\mathrm{A}}$ & $\mathsf{S}_2\mathsf{C}_{\mathrm{B}}$ & $\mathsf{S}_2\mathsf{C}_{\mathrm{C}}$ & $\mathsf{S}_2\mathsf{C}_{\mathrm{D}}$ \\ \hline
\multicolumn{2}{c|}{\multirow{2}{*}{Data}}   & $43.17$                       & $35.02$                 & $35.87$                      & $31.92$                  \\
&                                                          & \scriptsize $\pm 0.93$     & \scriptsize $\pm 3.14$    & \scriptsize $\pm 2.99$    & \scriptsize $\pm 1.33$\\
\hline
\multicolumn{2}{c|}{\multirow{2}{*}{Oracle}}   & $48.89$                       & $44.60$                & $44.48$                     & $42.27$                 \\
&                                                          & \scriptsize $\pm 0.97$     & \scriptsize $\pm 1.98$    & \scriptsize $\pm 1.68$    & \scriptsize $\pm 1.52$\\
\hline                                                               
\multicolumn{1}{c|}{\multirow{8}{*}{\rotatebox{90}{RFI-DRUNet} }}   & \multirow{2}{*}{$\mathsf{S}_2\mathsf{C}_{\mathrm{A}}$}   & 59.43        & 36.38      & 37.77       & 32.89 \\
\multicolumn{1}{c|}{} &                                                         & \scriptsize $\pm6.54$    & \scriptsize $\pm4.88$   & \scriptsize $\pm4.27$   & \scriptsize $\pm1.69$\\
\cline{2-6}
\multicolumn{1}{c|}{} & \multirow{2}{*}{$\mathsf{S}_2\mathsf{C}_{\mathrm{B}}$}   & 60.15        & 59.81       & 59.88       & 59.22 \\
\multicolumn{1}{c|}{} &                                                         & \scriptsize $\pm5.08$    & \scriptsize $\pm4.91$   & \scriptsize $\pm5.02$   & \scriptsize $\pm4.73$\\
\cline{2-6}                                                             
\multicolumn{1}{c|}{} & \multirow{2}{*}{$\mathsf{S}_2\mathsf{C}_{\mathrm{C}}$}   & 60.36        & 60.01       & 60.12       & 59.51 \\
\multicolumn{1}{c|}{} &                                                         & \scriptsize $\pm4.34$    & \scriptsize $\pm4.40$   & \scriptsize $\pm4.27$   & \scriptsize $\pm4.25$\\
\cline{2-6}                                                             
\multicolumn{1}{c|}{} & \multirow{2}{*}{$\mathsf{S}_2\mathsf{C}_{\mathrm{D}}$}   & 60.41        & 60.18       & 60.19      & 59.95 \\ 
\multicolumn{1}{c|}{} &                                                         & \scriptsize $\pm4.49$    & \scriptsize $\pm4.55$   & \scriptsize $\pm4.47$   & \scriptsize $\pm4.53$\\ \hline
\end{tabular}
\caption{{Scenario 2: restoration performance in terms of average PSNR and standard deviations computed over the test data sets.}}\label{tb:psnr_s2}
\end{table}

\begin{figure}[ht!]
\centering
\includegraphics[width=0.9\columnwidth]{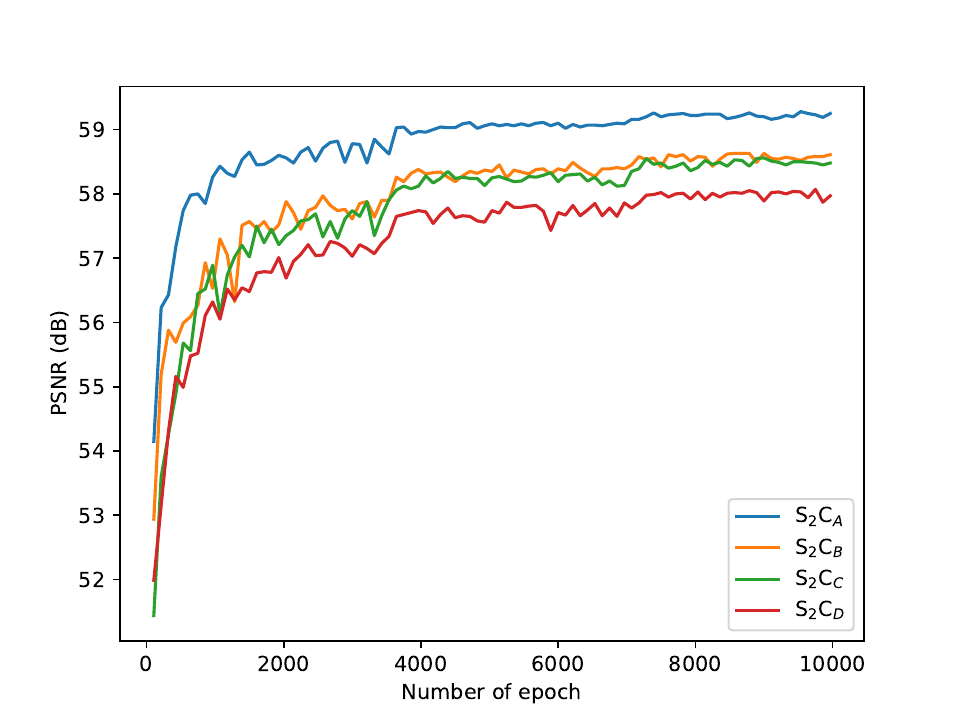}
\caption{Validation: restoration performance (in terms of PSNR) as a function of the number of epochs.}
\label{fig:val_psnr}
\end{figure}

\noindent \emph{Validation --} The restoration performance (in terms of PSNR) as a function of the number of epochs during the validation stage is depicted in Fig.~\ref{fig:val_psnr}. This figure shows that the validation results are consistent with the results obtained during the testing stages. The model $\mathsf{S}_2\mathsf{C}_{\mathrm{A}}$ reaches a higher validation result compared to the other models since it uses less variety of {RFI} during the training stage. Conversely, the models $\mathsf{S}_2\mathsf{C}_{\mathrm{B}}$ and $\mathsf{S}_2\mathsf{C}_{\mathrm{C}}$ have almost identical validation results, and the model $\mathsf{S}_2\mathsf{C}_{\mathrm{D}}$ shows slightly lower validation results due to the fact that it has to handle with all types of {RFI}. 
In the sequel of the paper, for brevity and unless otherwise specified, only the model $\mathsf{S}_2\mathsf{C}_{\mathrm{D}}$, which is expected to be more robust to any type of {RFI}, will be considered.\\

\noindent {\emph{Out-of-distribution data sets --}
To assess the robustness of the proposed RFI-DRUNet to experimental conditions that go beyond those encountered during the training stage, we now investigate the generalization ability of the model when analyzing out-of-distribution (OOD) data. This model mismatch is envisioned with respect to two aspects. First, the performance of the model  $\mathsf{S}_2\mathsf{C}_{\mathrm{D}}$, trained on dynamic spectra corrupted by an instrumental noise of $\textrm{SNR}\approx-5\textrm{dB}$ ($\sigma^2_{\mathrm{E}}=1$), is evaluated when restoring data corrupted by noises of higher levels, i.e., $\textrm{SNR}\in \left\{-8.6,-10.4,-12.6,-15.6\right\}[\textrm{dB}]$. Second, since RFI emitters may produce interference of more complex temporal and spectral profiles than those prescribed in Section \ref{subsec:simulation_RFI}, instances of RFI with sinusoidal shapes are also considered. More precisely, instead of generating RFI signals with the Gaussian shapes described in Eq. \eqref{eq:model_RFI}, nbct and/or bbt RFI signals are generated according to an oscillating profile and included into the testing sets. These additional RFI profiles are denoted nbct-sin and bbt-sin, depending on their temporal and spectral spreading. These complementary experiments aims at enriching the diversity of interference patterns and the range of noise levels during testing. Note that the model $\mathsf{S}_2\mathsf{C}_{\mathrm{D}}$ evaluated in what follows has not be re-trained on these newly generated data sets. Thus this experimental protocol ensures a more thorough assessment of the model adaptability to unforeseen interference conditions.  Table \ref{tb:mismatch_ood_test} reports the restoration results of the model when faced to these OOD data sets. It clearly appears that the restoration performance is weakly impacted by the noise level, up to a reasonable mismatch. Similarly, the presence of sinusoidal shaped RFI in the test sets barely affect the restoration performance, which confirms the robustness of the proposed method.}

\begin{table}\small
 \renewcommand{\arraystretch}{1.1}
\setlength{\tabcolsep}{6pt}
\centering
\begin{tabular}{c  c c c}
\hline 
& \multicolumn{ 2}{c}{\textbf{Simulation parameters}}    & {\textbf{PSNR}} \\ 
\hline
\multicolumn{ 1}{c|}{\multirow{5}{*}{{\makecell[c]{OOD \\ with respect to \\ noise level  }}}} &$\sigma^2_{\mathrm{E}}=1$  &   $\mathrm{SNR}=-5.6\mathrm{dB}$              &  $59.95$ \scriptsize$\pm 4.53$         \\
\cline{2-4}                                     
\multicolumn{ 1}{c|}{}& $\sigma^2_{\mathrm{E}}=2$  &   $\mathrm{SNR}=-8.6\mathrm{dB}$                &     $59.24$ \scriptsize$\pm0.81$          \\
\cline{2-4}  
\multicolumn{ 1}{c|}{}& $\sigma^2_{\mathrm{E}}=3$   &  $\mathrm{SNR}=-10.4\mathrm{dB}$                &       $57.81$ \scriptsize$\pm0.61$        \\
\cline{2-4}  
\multicolumn{ 1}{c|}{}& $\sigma^2_{\mathrm{E}}=5$   &  $\mathrm{SNR}=-12.6\mathrm{dB}$                &       $54.86$ \scriptsize$\pm1.56$        \\
\cline{2-4}  
\multicolumn{ 1}{c|}{}& $\sigma^2_{\mathrm{E}}=10$   &  $\mathrm{SNR}=-15.6\mathrm{dB}$                &       $50.84$ \scriptsize$\pm2.15$        \\
\hline
\hline
\multicolumn{ 1}{c|}{\multirow{3}{*}{\makecell[c]{OOD \\ with respect to \\ RFI profile} }} &\multicolumn{2}{c}{nbct-sin RFI}                   &  $59.90$ \scriptsize$\pm3.22$        \\
\cline{2-4}                                     
\multicolumn{ 1}{c|}{}& \multicolumn{ 2}{c }{bbt-sin RFI}                  &   $59.91$ \scriptsize$\pm3.17$         \\
\cline{2-4}  
\multicolumn{ 1}{c|}{}&\multicolumn{ 2}{c }{nbct-sin RFI + bbt-sin RFI }   &   $59.59$ \scriptsize$\pm3.11$        \\
\hline
\end{tabular}
\caption{{Out-of-distribution data sets: restoration performance in terms of average PSNR and standard deviations over the test data sets.} }
\label{tb:mismatch_ood_test}
\end{table}

\subsection{Detection results}
\label{Comparison with classification methods}
The proposed method is now simplified to turn it into a simple RFI detector whose performance can be compared to those reached by competitive methods from the literature. To do so, a binary mask $ \hat{M}(\cdot,\cdot)$ deciding the presence of possible RFI can be easily computed as
\begin{equation*}
    \hat{M}(\indFreq,\indTime) = \begin{cases}
                $1$, & \text{if $|\hat{S}(\indFreq,\indTime)-{S}(\indFreq,\indTime)| > \eta$}\\
                $0$, & \text{otherwise}
            \end{cases}
            \begin{tabular}{c}
                (presence of RFI)\\
                (absence of RFI)
            \end{tabular}
\end{equation*}
where $\hat{S}(\cdot,\cdot)$ and ${S}(\cdot,\cdot)$ are the restored dynamic spectrum provided by RFI-DRUNet and the input dynamic spectrum, respectively, and $\eta$ is a threshold balancing the probability of detection and the probability of false alarm. {This threshold has been fixed to $\eta=0.15$ such that it maximizes the F1-score of RFI-DRUNet to ease the comparisons to the other methods, as also done by \cite{kerrigan2019optimizing} and \cite{mesarcik2022learning}}. In particular, the performance of the proposed method is compared to the detection ability of two alternative deep learning-based RFI mitigation methods. The first considered network architecture is a U-Net, initially proposed to perform medical image segmentation tasks and later adapted for RFI mitigation tasks \citep{akeret2017radio}. The second compared model is RFI-Net \citep{yang2020deep}, based on an encoder-decoder architecture with residual blocks and batch normalization. {These two models have been trained on the data sets corresponding to scenario $\mathsf{S}_2\mathsf{C}_{\mathrm{D}}$. More precisely, during the training stage, the binary masks defined by Eq. \eqref{eq:RFI_masks} and randomly drawn during the data generation process are given as output labels to the models.}

As explained in Section \ref{Evaluation metric}, the methods are compared in terms of standard classification scores, namely precision, recall, F1-score, AUROC and AUPRC since they provide a comprehensive performance assessment of a binary classification model. These metrics are reported in Table \ref{tb:Comparison_results}. {Although the primary objective of the proposed RFI-DRUNet model is not to detect RFI but rather to restore dynamic spectra, these results demonstrate that its ability to flag RFI is comparable to those of state-of-the-art methods specifically designed to perform this task. It is worth noting that, contrary to RFI-DRUNet, U-Net and RFI-Net are unable to restore RFI-free dynamic spectra.}
  
\begin{table}\small
 \renewcommand{\arraystretch}{1.1}
\setlength{\tabcolsep}{4pt}
\centering
\begin{tabular}{l cccc cc}
\hline
\textbf{Method}    & \textbf{prec}  & \textbf{rec}      & \textbf{F1-score}         & \textbf{AUROC}   & \textbf{AUPRC}\\ \hline
\multirow{ 2}{*}{RFI-DRUNet}         & $0.972 $        & $0.961$   & $0.966$     &  $0.995$        &$0.986$               \\
                                     & \scriptsize $\pm 0.398$      & \scriptsize $\pm 0.0157 $  &  \scriptsize  $\pm 0.024$ &  \scriptsize  $\pm 0.002$                &  \scriptsize  $\pm 0.016$               \\
\hline                                     
\multirow{ 2}{*}{U-Net}              & $0.858$         & $0.987$          &  $0.917$  & $0.991$             & $0.926$               \\
                                     & \scriptsize  $\pm 0.048$     & \scriptsize  $\pm 0.005$        & \scriptsize  $\pm 0.029$  & \scriptsize  $\pm 0.003$            & \scriptsize  $\pm 0.027$              \\
\hline                                     
\multirow{ 2}{*}{RFI-Net}            & $0.864$     & {$0.988$ }  & $0.921$  & $0.987$             & $0.926$              \\ 
                                     & \scriptsize $\pm 0.047$     & \scriptsize {$\pm 0.005$ }  & \scriptsize  $\pm 0.029$  &\scriptsize  $\pm 0.004$             &\scriptsize  $\pm 0.026$       \\ 
\hline
\end{tabular}
\caption{{Detection performance of compared algorithms in terms of precision, recall, F1-score, AUROC and AUPRC. The results are reported with mean and standard deviation computed over the test data sets.}}
\label{tb:Comparison_results}
\end{table}

\begin{figure}
\centering
\includegraphics[width=0.95\columnwidth]{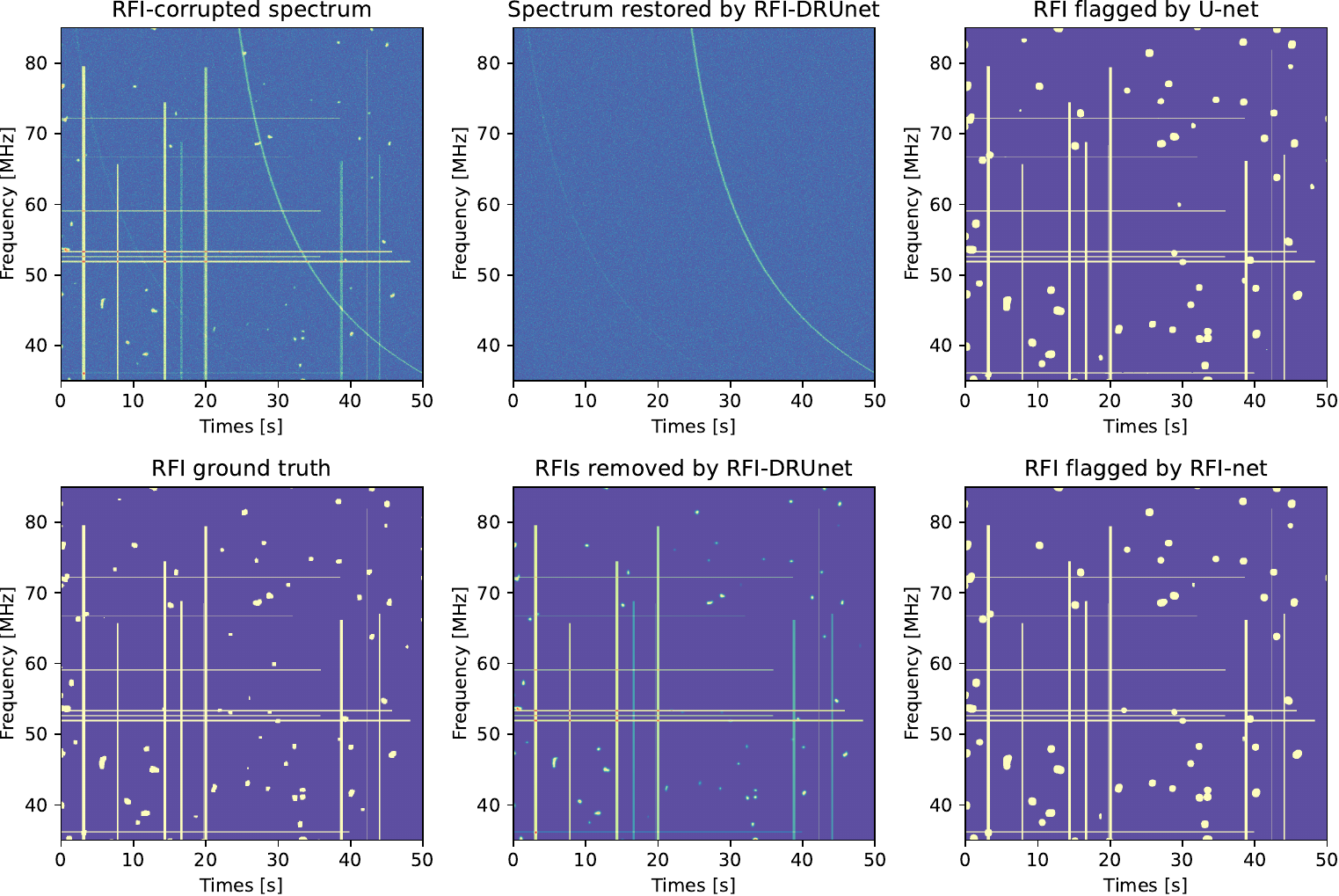} 
\caption{Visual comparison of the results for simulated spectrum from $\mathsf{S}_2\mathsf{C}_{\mathrm{D}}$ provided by the compared methods.}\label{fig:visual_comparison}
\end{figure}

To conduct a qualitative comparison of the results, outputs provided by the compared methods are depicted in Fig.~\ref{fig:visual_comparison}, as well as the spectrum restored and the {RFI} identified by RFI-DRUNet, for a particular signal.

\subsection{Illustration on a real observation}
\label{Illustration on a real data}
\begin{figure}
\centering
\includegraphics[width=0.95\columnwidth]{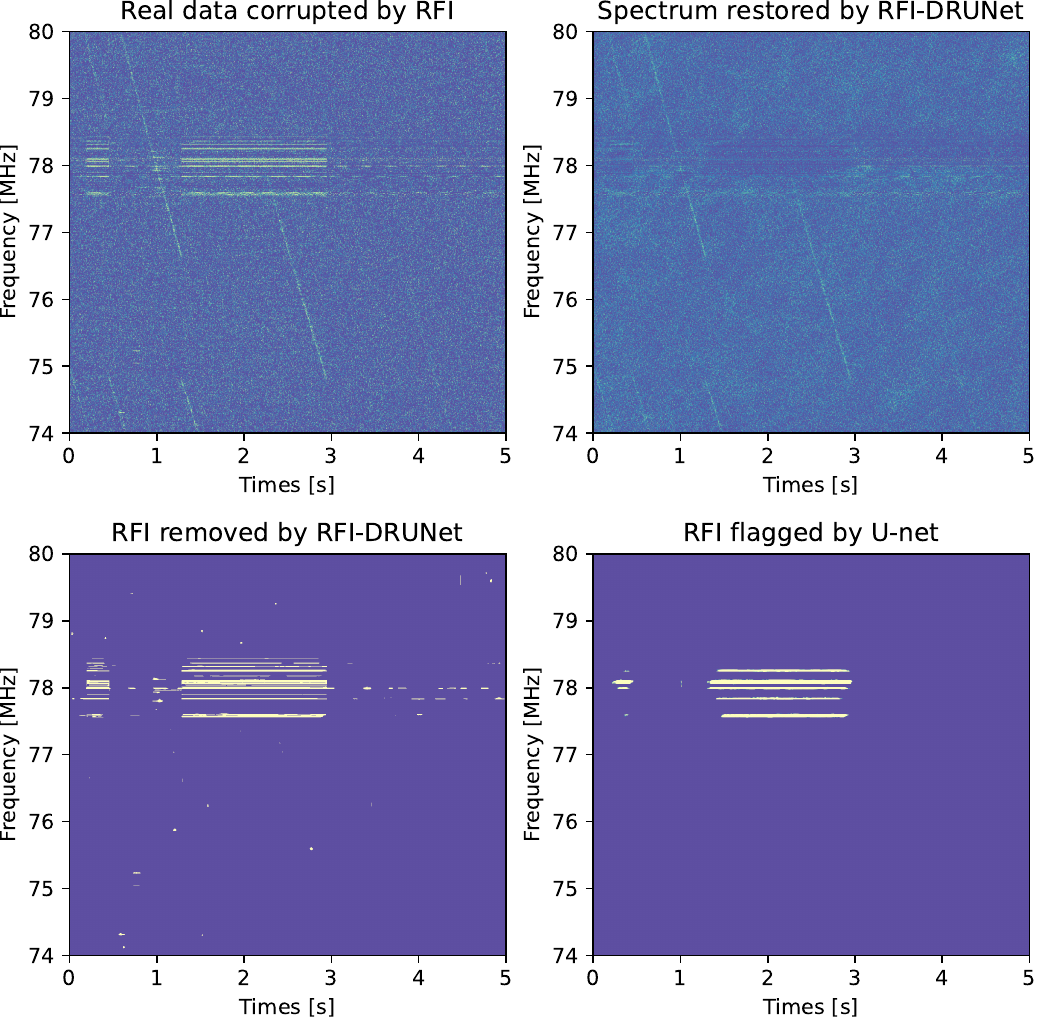}
\caption{Visual comparison of the results for a real data set provided by RFI-DRUNet and U-Net.}\label{fig:visual_real}
\end{figure}

Since the proposed method has already demonstrated outstanding restoration and detection performance on the simulated data sets, we now illustrate this performance on a real observation data from the NenuFAR telescope. A dynamic spectrum over the spectral range $74-80$MHz of duration of $5$s extracted from an observation for pulsar B1919+21 is considered, as depicted in Fig.~\ref{fig:visual_real} (top left panel).  {The compared models RFI-DRUNet, RFI-DRUnet and U-Net have been trained as in Section \ref{Comparison with classification methods}, i.e., using the synthetic data set $\mathsf{S}_2\mathsf{C}_{\mathrm{D}}$ generated following the protocol described in Section \ref{Synthetic Datasets}. This experiment somehow challenges the  generalization ability of the compared models. Indeed here they are tested on a real signal whose content is expected to substantially depart from the simplifying modeling assumptions underlying the simulation framework described in Section \ref{simulation framework}.} Figure \ref{fig:visual_real} (top right panel) shows the restored dynamic spectrum achieved by RFI-DRUNet, as well as a comparison results of the {RFI} detected by RFI-DRUNet and U-Net (bottom panels). {The results provided by RFI-Net are not reproduced in this manuscript since they are not of sufficient quality to be informative. The poor results exhibited by RFI-Net may be explained by its weak generalization ability.} In these panels, {RFI} mostly appear as a form of {nbct} corruptions around $78$MHz and persist from $0.2$s to $1.5$s. Both RFI-DRUnet and U-Net are proficient in detecting most of the {RFI}, but clearly, RFI-DRUNet provides higher accuracy, especially for the RFI around $78$MHz between $0.2$s and $1$s. RFI-DRUNet also detects several instances of pulse-like {RFI}, while U-Net hardly detects any. This result confirms the ability of RFI-DRUNet to detect RFI with accuracy, whatever their shapes, and also to restore the corrupted dynamic spectra while preserving most of the signal of interest.

\subsection{Discussion}
\label{Discussion}
\noindent \emph{Restoration vs.~detection --}
As stated earlier, it is worth noting that U-Net and RFI-Net have been designed to achieve an RFI-flagging task, i.e., to identify and locate the time-frequency bins possibly affected by {RFI} in the dynamic spectra. {The experimental results demonstrate exceptionally good performance achieved by these methods, which can be explained by the well-documented ability of convolutional neural networks to extract relevant features from 2-dimensional data, i.e., images in general and dynamic spectra in particular.} Once  {RFI} signals have been identified, several strategies can be envisioned to handle them in a subsequent pipeline of astronomical data processing. The most common strategy consists of throwing away the measurements corresponding to the pulsar period corrupted by {interference}. Hence these pixel-level segmentation methods appear to be suboptimal since they conduct to the loss of significant information. Conversely, the proposed RFI-DRUNet method is able not only to correctly flag the {RFI} but also to restore the corrupted time-frequency bins with reliable signal values. 

{To illustrate this loss of information and also the relevance of restoring plausible values, one considers the strategy that would consist in replacing time-frequency bins corrupted by RFI by plausible values. Such a strategy has been already considered in Section \ref{Denoising results} while reporting the restoration performance of the so-called oracle, assumed to perfectly know the RFI locations. Hereafter, the analysis goes one step further by considering realistic detectors, namely U-Net and RFI-Net, in addition to the oracle detector. In the noisy scenario ($\mathsf{S}_2\mathsf{C}_{\mathrm{D}}$), these bins are filled with random values drawn according to the instrumental noise statistical model, $\mathcal{N}(0,\sigma^2_{\mathrm{E}})$. Table \ref{tb:detection_restoration} reports the restoration results and also recalls the performance reached by the proposed RFI-DRUNet model. These results show that there is a significant gap in the quality of the data between the compared strategies.}

\begin{table}\small
 \renewcommand{\arraystretch}{1.15}
\setlength{\tabcolsep}{8pt}
\centering
\begin{tabular}{c  c c c}
\hline 
& \multicolumn{ 2}{c }{\textbf{Method}}       & \multirow{2}{*}{\textbf{PSNR}} \\ 
& {\textbf{Detection}}  & \textbf{Filling value}       &  \\ 
\hline
\multicolumn{ 1}{c|}{\multirow{5}{*}{\rotatebox{90}{$\mathsf{S}_2\mathsf{C}_{\mathrm{D}}$} }} &\multicolumn{ 2}{c }{Data}                   &  31.91 \scriptsize$\pm$1.33         \\
\cline{2-4}                                     
\multicolumn{ 1}{c|}{}& \multicolumn{ 2}{c }{RFI-DRUNet}                 &  $59.95$ \scriptsize$\pm4.53$          \\
\cline{2-4}  
\multicolumn{ 1}{c|}{}& Oracle            &  $\mathcal{N}(0,\sigma_{\mathrm{E}}^2) $           &   $42.27$ \scriptsize$\pm1.52$        \\
\cline{2-4}  
\multicolumn{ 1}{c|}{}&U-Net             & $\mathcal{N}(0,\sigma_{\mathrm{E}}^2) $           &   $41.64$ \scriptsize$\pm1.33 $        \\
\cline{2-4}  
\multicolumn{ 1}{c|}{}&RFI-Net           & $\mathcal{N}(0,\sigma_{\mathrm{E}}^2) $           &    $41.69$ \scriptsize$\pm 1.32$         \\
\hline
\end{tabular}
\caption{{Restoration performance of compared algorithms in terms of average PSNR and standard deviation computed over the test data sets.}}
\label{tb:detection_restoration}
\end{table}

{Interestingly, while it has not been specifically designed to perform  this task, RFI-DRUNet also reaches results comparable to those obtained by U-Net and RFI-net when conducting RFI detection/flagging.} These significant improvements may come from several technical aspects adopted in this work. First, the architecture of RFI-DRUNet is different from those of the compared methods. Inherited from DRUNet, it is known to provide better results when facing a denoising task. {This improvement does not systematically result from a higher complexity since RFI-DRUNet embeds a smaller number of network parameters to be adjusted when compared to the RFI-Net, which significantly reduces the computational burden during the training and testing stages. To illustrate, the number of parameters defining the compared models are reported in Table \ref{tab:number_parameters}.} Second, the pretext (denoising-like) task chosen to design the training loss function minimized by RFI-DRUNet is much more demanding than the ones adopted by U-Net and RFI-Net. When finally {simplified}, the proposed method is able to solve a easier task. Conversely, U-Net and RFI-Net may reach even better RFI flagging performance after adapting their architecture and their loss functions to be trained on a restoration task. In other words, the capacity of those models may be under-exploited by a too simple training strategy.\\

\begin{table}[ht!]\small
 \renewcommand{\arraystretch}{1.15}
\setlength{\tabcolsep}{8pt}
\centering
\begin{tabular}{c  c c c}
\hline 
       & {\textbf{RFI-DRUNet}} & {\textbf{RFI-Net}}  & \textbf{U-Net}       \\ 
       \hline
{\makecell[c]{ $\sharp $ parameters \\ ($\times 10^6$)  }} &   32.65     & 48.21  & 17.26\\
\hline
\end{tabular}
\caption{{Number of parameters of the compared models.}}
\label{tab:number_parameters}
\end{table}

\begin{figure}[!htbp]
\centering
\includegraphics[width=0.95\columnwidth]{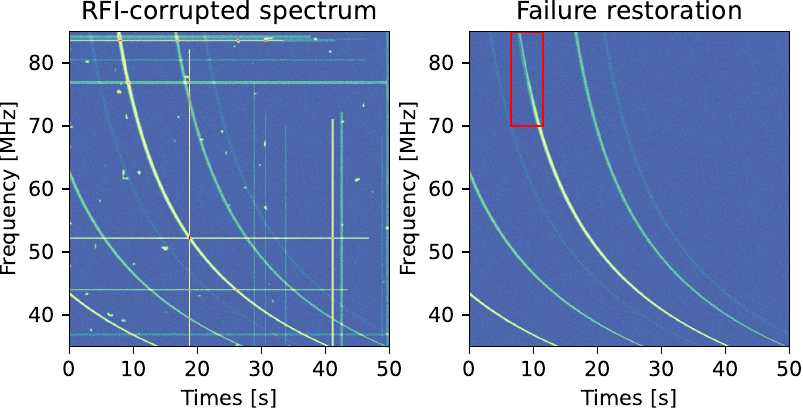}
\caption{Example of a failure situation from data set corresponding to scenario $\mathsf{S}_2\mathsf{C}_\mathrm{D}$. The amplitude of the part of the pulsar signal (framed in the red box) has been incorrectly reduced by RFI-DRUNet.}\label{fig:bad_prediction}
\end{figure}

\noindent \emph{Failure situation --}
To provide a fair analysis of the potential brought by the proposed method, one finally wants to point out the fact that RFI-DRUNet may behave quite poorly in some particular situations. Indeed, a careful (empirical) inspection of the restoration results obtained by the network shows that some restoration scores are significantly lower than the overall score averaged over the whole data set. Figure \ref{fig:bad_prediction} shows one archetypal example of failure. It clearly appears that the method erroneously treats the part of the pulsar signal as a possible RFI signal and tends to decrease its amplitude. This phenomenon typically arises at high frequencies where the pulsar signal is slightly affected by dispersion. The pulsar signal is then almost vertical and has a shape that is very similar to {nbt} {RFI}, i.e., temporally localized but spectrally spread {interference}.

\section{Application to estimation of pulsar TOAs}
\label{Estimaion of TOA}
Pulsars have a highly stable period of rotation, so the signals received from their emissions have a consistent period as well. The study of pulsars by precisely measuring their so-called times-of-arrival (TOAs) is known as chronometry. By investigating the pulsar TOAs, one can deduce various physical parameters intrinsic to pulsars and other astronomical quantities associated with various problems such as cosmic clocks and the detection of gravitational waves. In what follows, we show that, thanks to its ability to detect {RFI} and restore dynamic spectra, the proposed RFI-DRUNet method is able to contribute to the improvement of the TOA estimation.

To estimate the pulsar TOA, the profile of a pulsar observed over a short time observation interval is generally assumed to deviate only slightly from the shape computed by integrating over a long period of time, hereafter referred to as the pulsar template. Leveraging this assumption, the observed signal pulsar $P(\indFreq,\indTime)$ can be modeled as a noisy, scaled, and temporally shifted version of the template signal denoted $\bar{P}(\indFreq,\indTime)$, i.e., 
\begin{align*}
    P(\indFreq,\indTime) = u + v \bar{P}(\indFreq,\indTime-\Delta t)+E(\indFreq,\indTime)
\end{align*}
where $u$ is an amplitude offset, $b$ is a scaling factor, $\Delta t$ is the temporal shift between the observed signal and the template, and $E(\indTime,\indFreq)$ is a term accounting for modeling errors. \cite{taylor1992pulsar} introduced a least squares method to estimate the offset $\Delta t$  between the observed signal and the template signal. The optimization problem is formulated in the Fourier domain. It provides an estimation with an accuracy of $0.1\delta t$, which is better than what can be obtained in the time domain. To illustrate the relevance of the proposed RFI-DRUNet method, we consider synthetic signals generated according to the scenario $\mathsf{S}_2\mathsf{C}_{\mathrm{D}}$ with a pulsar template $\bar{P}(\cdot,\cdot)$ fully described by Eq.~\eqref{eq:pulsar_profile} and a pulsar period fixed to $\rho = 64$ [bins]. Then Taylor's estimation method was applied to the RFI-free signal, the RFI-corrupted signal and the signal restored by RFI-DRUNet. In this controlled experimental setup, the true value of $\Delta t$ is perfectly known. Thus the accuracy of the estimation conducted on the $3$ types of signals can be qualitatively assessed by computing a mean square error (MSE) and a mean absolute deviation (MAE).

\begin{figure}[h!]
\centering
\includegraphics[width=0.85\columnwidth]{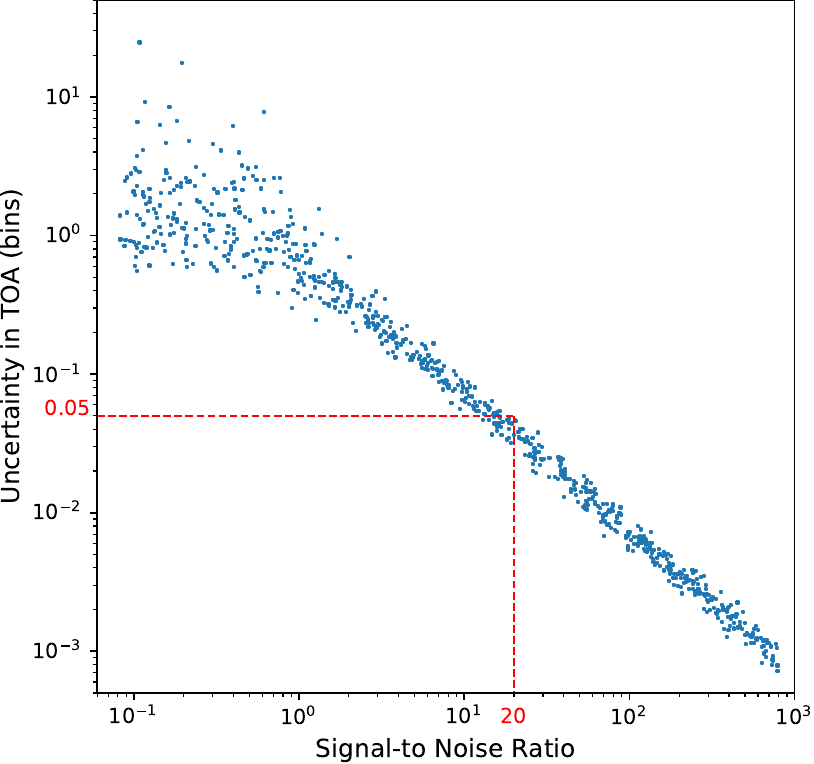}
\caption{Scatter plot of the TOA estimation uncertainty vs. SNR. The red dashed lines point out the maximum SNR value which is  $20$, considered in Scenario $\mathsf{S}_2\mathsf{C}_{\mathrm{D}}$ and the corresponding uncertainty value adjusted by empirical linear regression (in the log-log space). }
\label{fig:snr-sigmaR}
\end{figure}

It is worth noting that Taylor's method grants its TOA estimation with an uncertainty measure denoted $\sigma_{\Delta t}$. This uncertainty is related to the difficulty of the estimation task and, for instance, is driven by the SNR level: lower the SNR, higher the uncertainty. This finding is illustrated in Fig.~\ref{fig:snr-sigmaR} where each dot is associated to a particular signal generated according to the scenario $\mathsf{S}_2\mathsf{C}_{\mathrm{D}}$ for a wider range of SNR (from $10^{-1}$ to $10^3$). Each signal is characterized by its SNR (x-axis) and the uncertainty measure provided by the estimation method (y-axis). In this log-log scatter plot, the uncertainty is clearly (inversely) proportional to the SNR. As reported in Section \ref{subsec:Simulation parameters}, the maximum  SNR value in the generated data set $\mathsf{S}_2\mathsf{C}_{\mathrm{D}}$ is $20$, which empirically corresponds to an uncertainty measure lower-bounded by $\sigma_{\Delta t}=0.05$. This complementary information provided by Taylor's method can be used to throw away all estimates accompanied by an uncertainty measure higher than a given threshold since they are considered as unreliable.

\begin{figure}[ht!]
\centering
\includegraphics[width=0.95\columnwidth]{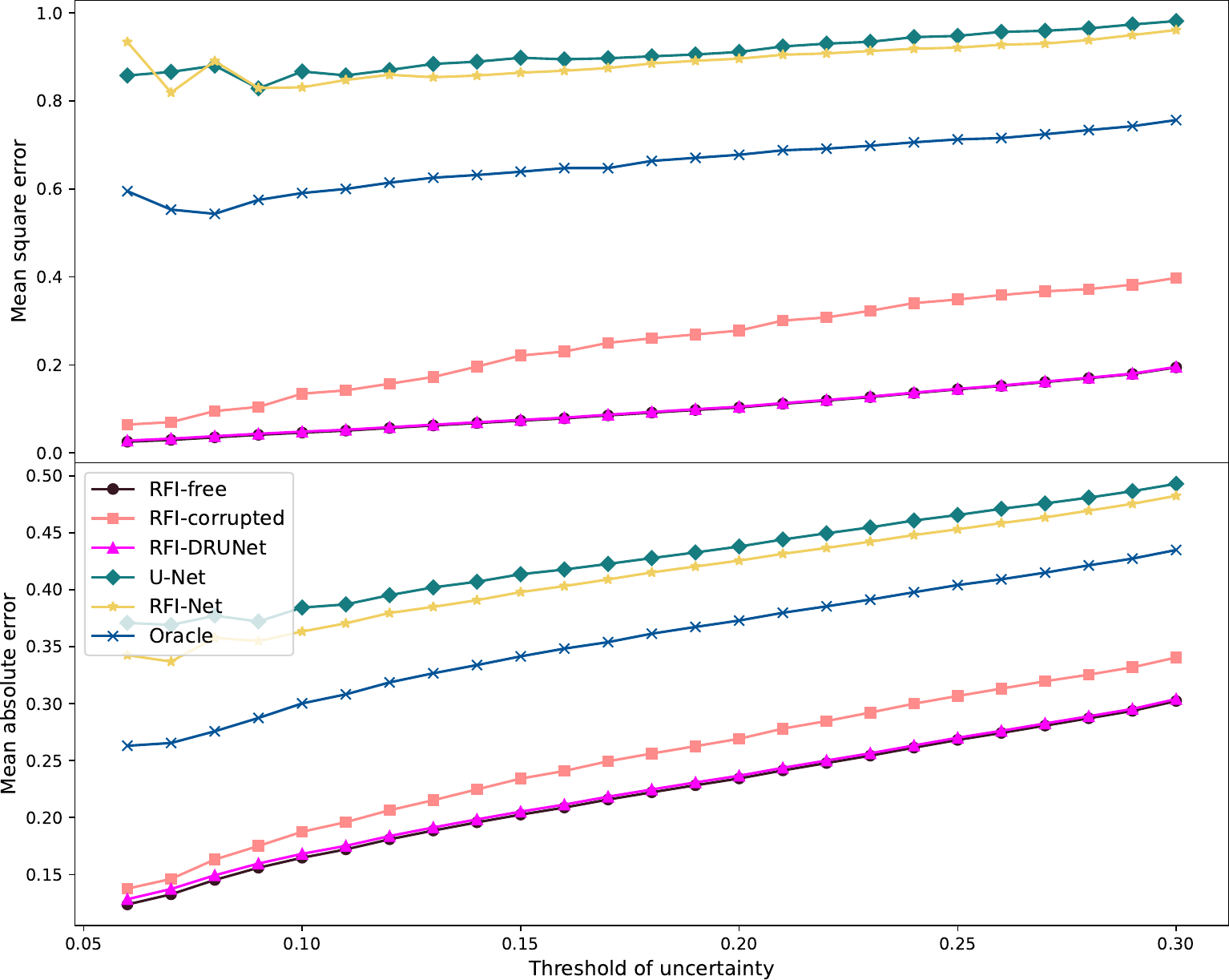}
\caption{TOA estimation errors in terms of MSE (top) and MAE (bottom) as functions of the uncertainty threshold obtained from RFI-free (blue line), RFI-corrupted (orange line) and restored (green line) dynamic spectra.}
\label{fig:toa_res}
\end{figure}

{Figure \ref{fig:toa_res} depicts the estimation errors (in terms of MSE and MAE) as functions of the uncertainty threshold. They have been obtained when the Taylor's method has been applied to the RFI-free signals, the RFI-corrupted signals and the signals restored by RFI-DRUNet. As a follow-up of the discussion in Section \ref{Discussion}, the Taylor’s method has been applied on the signals retrieved by three alternative approaches: RFI signals are located/flagged by the oracle detector, U-Net or RFI-Net, and the time-frequency bins identified as corrupted are replaced by random values.} As expected, the lowest errors are obtained when estimating the TOAs from the RFI-free dynamic spectra. When the estimation is conducted on RFI-corrupted dynamic spectra, it yields a large discrepancy with the theoretical TOA, which is quite understandable since the pulsar signal is exposed to RFI contamination. The errors obtained from the dynamic spectra restored by RFI-DRUNet are close to those obtained by the RFI-free spectrum. This confirms that the proposed method significantly improves the estimation of the pulsar TOA when RFI-corrupted dynamic spectra have been restored beforehand. {U-Net and RFI-Net both exhibited errors higher than those of the the oracle method. It is also important to highlight that the three alternative methods result in significantly larger errors compared to the TOA estimations derived from the RFI-corrupted signal. Again, this finding supports the idea that a substantial loss of information occurs when flagging and replacing corrupted bins with random values. On the contrary, the proposed RFI-DRUNet restoration model consistently provides reliable estimations.}

\section{Summary and conclusions}
\label{sec:conclusion}
This paper formulated the problem of radio frequency interference mitigation (RFI) as a restoration task, to go beyond conventional approaches which aimed only at detecting and localizing RFI in dynamic spectra. {To conduct this task, a new deep neural network, coined as RFI-DRUNet, was designed by leveraging and customizing a popular network proposed in the computer vision literature. To train this new model, a whole framework was designed to produce simulated RFI-free and corresponding RFI-corrupted dynamic spectra}. This framework was instantiated in the specific context of pulsar observations and relied on physics-inspired and statistical models of the pulsar signals and of the {RFI}. The relevance of the approach adopted in this paper was assessed thanks to an extensive set of numerical experiments which demonstrated the ability of RFI-DRUNet not only to identify RFI but also to restore the corrupted dynamic spectra. The interest of the method was illustrated by monitoring the expected gain in the accuracy reached when estimating pulsar time-of-arrivals. {Future works include taking the phase information into account during the  restoration process. They  will be also devoted to the compression of the deep network to reduce its computational complexity, with the aim of its integration into a real-time processing chain.}

\section*{Acknowledgements}
The Nan\c{c}ay Radio Observatory is operated by the Paris Observatory, associated with the French Centre National de la Recherche Scientifique (CNRS), and partially supported by the R\'egion Centre, France.
The development of NenuFAR has been supported by staff and funding from Station de Radioastronomie de Nan\c{c}ay, CNRS-INSU, Observatoire de Paris-PSL, Université d’Orléans, Observatoire des Sciences de l’Univers en Région Centre, Région Centre-Val de Loire, DIM-ACAV and DIM-ACAV+ of Région Ile-de-France, Agence Nationale de la Recherche. The authors warmly acknowledge Dr. Louis Bondonneau for his help to produce the plots in Figure \ref{fig:pulsar_profiles}.

\bibliographystyle{model2-names} 
\bibliography{citation}

\end{document}